\documentclass[twocolumn,trackchanges]{aastex7}
\usepackage{appendix}
\usepackage{graphicx}
\usepackage{CJKutf8}
\usepackage{longtable} 
\usepackage{booktabs} 
\usepackage{comment}
\usepackage{amsmath}
\usepackage{xcolor}

\newcommand{\Msun}{${\rm M}_{\odot}$}

\begin{document}
\begin{CJK}{UTF8}{gbsn}
\title{SN 2023axu: A Type IIP Supernova Interacted with a Low-Density Stellar Wind}

\author{Zeyi Wang}
\affiliation{Yunnan Observatories (YNAO), Chinese Academy of Sciences (CAS), Kunming, 650216, China}
\affiliation{International Centre of Supernovae (ICESUN), Yunnan Key Laboratory of Supernova Research, Kunming 650216}
\affiliation{School of Astronomy and Space Science, University of Chinese Academy of Sciences, Beijing, 101408, China}

\email{wangzeyi@ynao.ac.cn} 

\author[0000-0002-8296-2590]{Jujia Zhang}
\affiliation{Yunnan Observatories (YNAO), Chinese Academy of Sciences (CAS), Kunming, 650216, China}
\affiliation{International Centre of Supernovae (ICESUN), Yunnan Key Laboratory of Supernova Research, Kunming 650216}
\email[show]{jujia@ynao.ac.cn}

\author[0009-0002-3956-6143]{Qian Zhai}
\affiliation{Yunnan Observatories (YNAO), Chinese Academy of Sciences (CAS), Kunming, 650216, China}
\affiliation{International Centre of Supernovae (ICESUN), Yunnan Key Laboratory of Supernova Research, Kunming 650216}
\email{} 

\author[0009-0003-3758-0598]{Liping Li}
\affiliation{Yunnan Observatories (YNAO), Chinese Academy of Sciences (CAS), Kunming, 650216, China}
\affiliation{International Centre of Supernovae (ICESUN), Yunnan Key Laboratory of Supernova Research, Kunming 650216}
\email{} 

\author[0000-0002-3334-4585]{G. Valerin}
\affiliation{Osservatorio Astronomico di Padova, Vicolo dell'Osservatorio 5, 35122 Padova, Italy}
\email{giorgio.valerin@inaf.it} 

\author[0000-0003-4254-2724]{A. Reguitti}
\affiliation{Osservatorio Astronomico di Padova, Vicolo dell'Osservatorio 5, 35122 Padova, Italy}
\email{andrea.reguitti@inaf.it} 

\author[0000-0002-7259-4624]{A. Pastorello}
\affiliation{Osservatorio Astronomico di Padova, Vicolo dell'Osservatorio 5, 35122 Padova, Italy}
\email{andrea.pastorello@inaf.it}

\author[0009-0009-2664-8212]{Zhenyu Wang}
\affiliation{Yunnan Observatories (YNAO), Chinese Academy of Sciences (CAS), Kunming, 650216, China}
\affiliation{International Centre of Supernovae (ICESUN), Yunnan Key Laboratory of Supernova Research, Kunming 650216}
\affiliation{School of Astronomy and Space Science, University of Chinese Academy of Sciences, Beijing, 101408, China}
\email{} 

\author[0009-0005-2963-7245]{Zeyi Zhao}
\affiliation{Yunnan Observatories (YNAO), Chinese Academy of Sciences (CAS), Kunming, 650216, China}
\affiliation{International Centre of Supernovae (ICESUN), Yunnan Key Laboratory of Supernova Research, Kunming 650216}
\affiliation{School of Astronomy and Space Science, University of Chinese Academy of Sciences, Beijing, 101408, China}
\email{}  

\author[0000-0002-3876-6330]{Tengfei Song}
\affiliation{Yunnan Observatories (YNAO), Chinese Academy of Sciences (CAS), Kunming, 650216, China}
\email{stf@ynao.ac.cn} 

\author[0000-0002-7714-493X]{Yongzhi Cai}
\affiliation{Yunnan Observatories (YNAO), Chinese Academy of Sciences (CAS), Kunming, 650216, China}
\affiliation{International Centre of Supernovae (ICESUN), Yunnan Key Laboratory of Supernova Research, Kunming 650216}
\affiliation{Osservatorio Astronomico di Padova, Vicolo dell'Osservatorio 5, 35122 Padova, Italy}
\email{957989597@qq.com} 

\begin{abstract}

We present photometric and spectroscopic observations of Type IIP supernova SN 2023axu, spanning $\sim$ 400 d after the explosion. Its light curve is typical of normal SNe IIP, with a $V$- band peak of −17.25 $\pm$ 0.06 mag and no early-time excess indicative of strong circumstellar interaction. The early spectra exhibit a distinctive broad ``ledge" near 4600 \AA. Through spectral modeling and comparison, we attribute this feature to a blend of C, N, and He lines excited by weak interaction between the ejecta and a low-density stellar wind. The late-time photometric evolution shows no discernible contribution from interaction, arguing against strong late-time circumstellar material engagement and supporting the low density wind scenario. From modeling, this SN synthesized $\sim 0.055\ \rm M_\odot$ of $^{56}$Ni, and nebular spectrum analysis indicates a progenitor mass near 15 $M_\odot$. SN 2023axu thus exemplifies weak ejecta–wind interaction and highlights the diversity of mass-loss histories and circumstellar environments of SNe II progenitor.

\end{abstract}


\section{\textbf{INTRODUCTION}} 
Core-collapse supernovae (CCSNe; \citealp{Heger_2003}) mark the explosive endpoints of massive stars ($M >$ 8 \Msun), serving as vital probes of their final evolutionary stages, including mass-loss history and circumstellar environments. As the most numerous class of CCSNe, Type II SNe (SNe II) retain a substantial hydrogen envelope at the time of explosion. The interaction between the expanding ejecta and pre-existing circumstellar material (CSM) is a critical process in SNe~II. It can be a dominant source of emission, particularly at early phases and serves as a unique diagnostic for the mass-loss history of the progenitor star \citep{2014ARA&A..52..487S}.

Observations reveal a broad continuum of CSM interaction strengths. For instance, SNe IIn exhibit long-lived narrow emission lines along with luminous, slowly declining light curves, indicating sustained interaction with dense, confined CSM (e.g., \citealp{2012AJ....144..131Z,2014Natur.509..471G,2020A&A...635A..39T}). In contrast, normal SNe IIP—characterized by an optical plateau powered by hydrogen recombination \citep{1993ApJ...414..712P}—typically show little evidence of ongoing CSM interaction. SNe IIL, which are defined by their light curves that decline almost linearly in magnitude, also fall into this category of generally weak interaction. However, a growing sample of early-time spectra has identified relatively short-lived flash-ionized features in some SNe IIP/IIL, implying the presence of confined CSM that is quickly overrun by the ejecta (e.g., \citealp{2017NatPh..13..510Y,10.1093/mnras/staa2273,Bruch_2021, Bruch_2023,2024ApJ...970..189J}).

High-cadence studies of nearby SNe II have recently revolutionized our view of this brief interaction phase. Landmark events such as SN 2023ixf and SN 2024ggi, for instance, have enabled the detailed capture of early shock breakout signals and the direct probing of dense CSM shed in the progenitor's final years (e.g., \citealp{2023ApJ...956L...5B,2023ApJ...956...46S,2024Natur.627..754L,2024Natur.627..759Z,2024ApJ...970L..18Z,2025SciA...11x2925Y}). These studies indicate that mass-loss rates can increase by orders of magnitude shortly before explosion, unveiling vigorous pre-SN activity in red supergiant progenitors (e.g., \citealp{2023SciBu..68.2548Z,2024ApJ...972L..15S,2024ApJ...969L..15X}).

Beyond these recognized signatures of clear but often short-lived interaction, there may exist SNe exhibiting even weaker and more fleeting CSM interaction. A potential spectroscopic indicator is a broad emission ``ledge” near 4600 \AA\ observed in the early-phase spectra of some SNe II (e.g., \citealp{Szalai_2019,Soumagnac_2020,10.1093/mnras/staf893}). The origin of this feature remains debated; proposed explanations include blending of metal lines (e.g., C, N, He) or a signature of weak interaction with a low-density, extended stellar wind \citep{Hosseinzadeh_2018, Bruch_2021}. If it indeed stems from weak CSM interaction, this ledge could provide a unique window into low mass-loss regimes that are difficult to detect through traditional narrow-line signatures.

SN 2023axu presents an excellent case to test this scenario. It displays the distinct ledge feature in its early spectra, while its overall light curve evolution closely follows that of normal SNe IIP, showing no signs of the significant early-time excess brightness indicative of strong CSM thermalization. This presents a clear dichotomy: a spectroscopic hint of interaction alongside photometric behavior consistent with a tenuous CSM environment. While the ledge feature has been analyzed by \cite{Shrestha_2024},  our study conducts an independent and comprehensive investigation based on extensive photometric and spectroscopic observations spanning approximately 1000 days. This dataset allows us to measure the key parameters of SN 2023axu, analyze the origin of the controversial ledge feature, and discuss its late-time evolution.  Our findings suggests that the properties of SN 2023axu are consistently explained by weak ejecta interaction with a low-density stellar wind, providing an example of a SN II progenitor with modest terminal mass loss. This work highlights the diversity of pre-SN mass-loss histories and underscores the importance of subtle spectroscopic signatures in uncovering them. 

The paper is structured as follows: Section \ref{sec:observations} describes the observations and data reduction. Section \ref{sec:analysis} presents the analysis of the light curves, bolometric luminosity, spectral evolution, and nickel mass. Section \ref{sec:discussion} discusses the physics of the ledge feature, the inferred progenitor and CSM properties, and the implications for our understanding of SNe II diversity. Section \ref{sec:conclusion} provides a summary of our conclusions.

\begin{figure}
\includegraphics[width=8.4cm,angle=0]{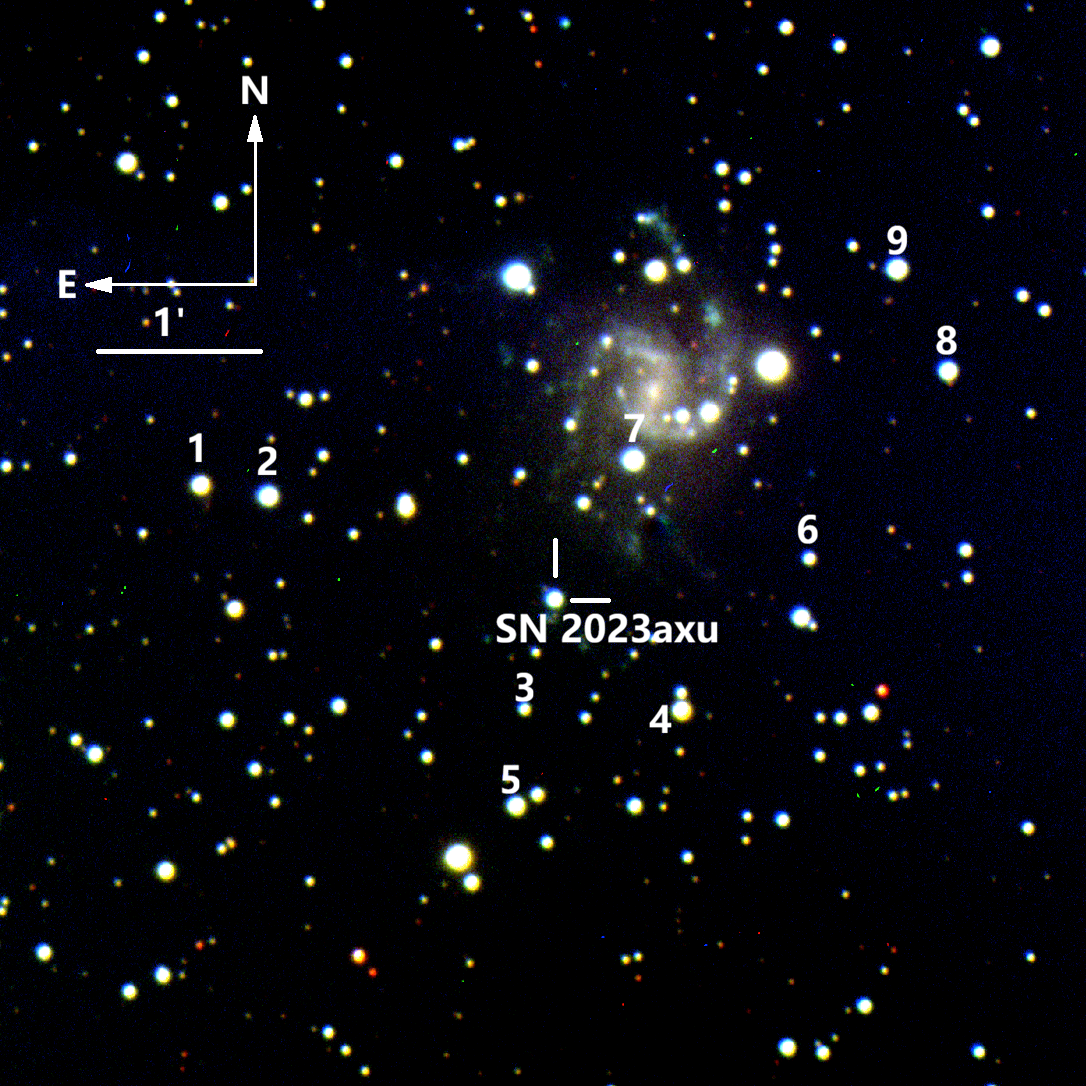}
\caption{Finder chart for SN 2023axu in the galaxy NGC~2283. The local reference stars listed in Table \ref{tab:finder} are marked with a nearby number. The image is a color composite created from g-, r-, and i-band observations obtained with the YFOSC instrument on the LJT.}
\label{fig:finder_chart}
\end{figure}

\begin{figure*}[ht!]
\plotone{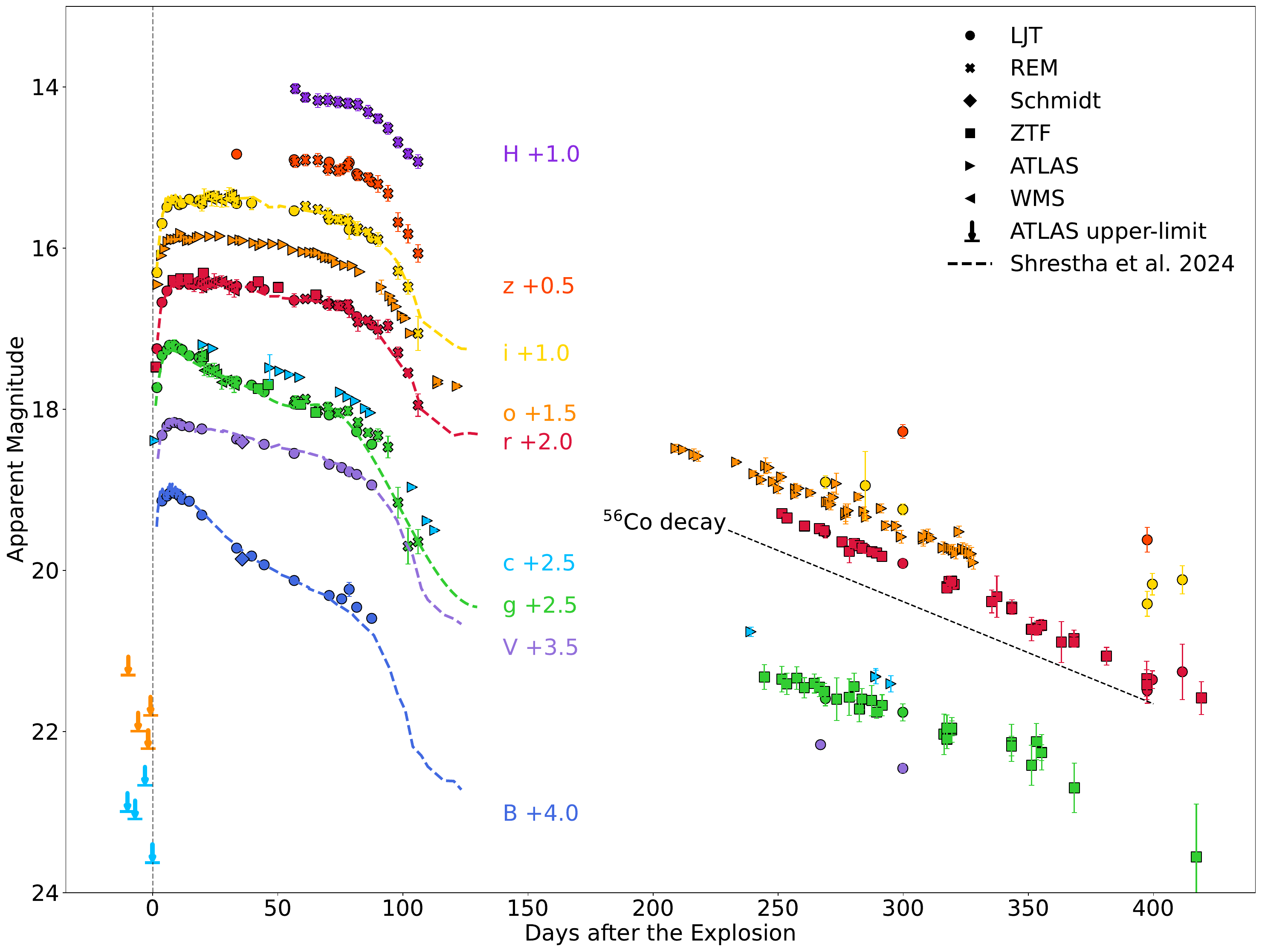}
\caption{Light curves of SN 2023axu. The different bands are vertically shifted for clarity. Data from various telescopes and sources are indicated accordingly. }
\label{fig:lightcurve}
\end{figure*}

\section{Observations} \label{sec:observations}

\subsection{Discovery}

SN 2023axu was discovered by the Distance Less Than 40 Mpc (DLT40) survey on January 28, 2023 (MJD 59972.11), in a clear-filter image with a magnitude of 15.64 ± 0.01 \citep{Sand_2023}. It is located at J2000 coordinates RA = {06$^{\rm h}$}{45$^{\rm m}$}{55$\fs$32}, Dec = {-18$\degr$}{13$\arcmin$}{53$\farcs$52}, in the SB(s)cd spiral galaxy NGC 2283 (Figure \ref{fig:finder_chart}), at a heliocentric redshift of z = 0.002805 $\pm$ 0.000005 \citep{Koribalski_2004}. \cite{Bostroem_2023tns} classified SN 2023axu as a young SN II based on its blue continuum and weak, broad hydrogen features. Ten hours later, an independent SN II classification was reported by \cite{Li_2023} based on observations with the 2.4 m Li-Jiang Telescope (LJT; \citealp{2015RAA....15..918F}) at the Li-Jiang Observatory of Yunnan Observatories. The LJT, equipped with the Yunnan Faint Object Spectrograph and Camera (YFOSC; \citealp{2019RAA....19..149W}), was subsequently employed for ongoing photometric and spectroscopic monitoring of this SN.

\subsection{Photometry}
Figure~\ref{fig:lightcurve} presents the optical and near-infrared light curves of SN 2023axu, covering a period of about 410~d, as obtained with LJT, Wumingshan 60cm Telescope (WMS) at Daocheng, the Rapid Eye Mount (REM) and Schmidt 67/92 telescopes. All images were processed following standard IRAF procedures, which included bias subtraction, flat-fielding, cosmic ray removal, and template subtraction. The point-spread-function (PSF) fitting method \citep{1987PASP...99..191S} was used to obtain instrumental magnitudes. These instrumental magnitudes were subsequently converted into standard Johnson $BV$ \citep{1966CoLPL...4...99J}, standard Sloan $griz$ (AB magnitude) and 2MASS H-band (Vega magnitude). For reference, data from \cite{Shrestha_2024} are overlaid, showing a light-curve evolution that matches our observations well. Additionally, we acquired publicly available data from both the ATLAS survey (c-band and o-band) and the ZTF survey (g-band and i-band), which cover the late stages of SN 2023axu.

Based on shock-cooling model fits, \cite{Shrestha_2024} estimated the explosion date of SN 2023axu to be MJD $ = 59971.48 \pm 0.03$. To independently constrain the explosion epoch, we applied a simple expanding fireball model to the early-time r-band light curve, expressed as: $F(t)=A\times(t-t_{0})^{2}+B$. The best fit yields coefficients of $A = 7.49 \times 10^{-16}$ and $B = 1.28 \times 10^{-15}$, resulting in $t_0 = 59971.5 \pm 0.5$. Considering the combined statistical uncertainty from the fireball model fit and the potential systematic offset between the model assumption and the actual explosion epoch, we adopt $59971.5 \pm 0.5$ as the explosion time of SN 2023axu. This result is consistent with the ATLAS non-detection on MJD = 59970.89 (c-band upper limit at $\sim$ 20.3 mag) and its first detection on MJD = 59971.91 (c-band magnitude of 15.93 $\pm$ 0.01 mag).

The archival Hubble Space Telescope (HST) observations of SN 2023axu (Program 17502; PI: David Thilker) were obtained using the Wide Field Camera 3 (WFC3/UVIS) on September 24, 2025 (MJD 60942.8), providing late-time photometry approximately 1000 days after the explosion. These data are publicly accessible through the Mikulski Archive for Space Telescopes (MAST) at STScI and are listed in Table \ref{Tab:HST}. We retrieved the Level 3 calibrated images from MAST, reduced with the version 12.1.4 of the HST Calibration Reference Data System (CRDS). We performed PSF photometry on the source identified at the coordinates of SN 2023axu with the \textsc{ecsnoopy}pipeline\footnote{\textsc{ecsnoopy} is a package for SN photometry using PSF fitting and/or template subtraction developed by E. Cappellaro. A package description can be found at http://sngroup.oapd.inaf.it/ecsnoopy.html.} v. 3.1.029, specifically adapted to run on images from space telescopes \citep[see][]{Reguitti2025A&A...698A.129R}. A forced PSF photometry is performed at the expected position of the transient. If no source is fitted with a flux higher than the decided threshold, then a limiting magnitude is provided with a flux equal to the selected threshold. Finally, the calibration of the instrumental magnitudes was done by calculating the photometric zero points in the AB scale with the formula\footnote{https://www.stsci.edu/hst/instrumentation/acs/data-analysis/zeropoints}:
\begin{align}
ZP_{AB} &= -2.5 \log_{10}(\text{PHOTFLAM}) \nonumber \\
    &\quad - 5 \log_{10}(\text{PHOTPLAM}) - 2.408
\end{align}
the PHOTFLAM and PHOTPLAM values were retrieved from the respective keywords stored in the header of the images.
In case the source is not identified (as in the $F275W$ and $F336W$ filters), we estimated a 3$\sigma$ upper limit.

\begin{figure}
\includegraphics[width=8.4cm,angle=0]{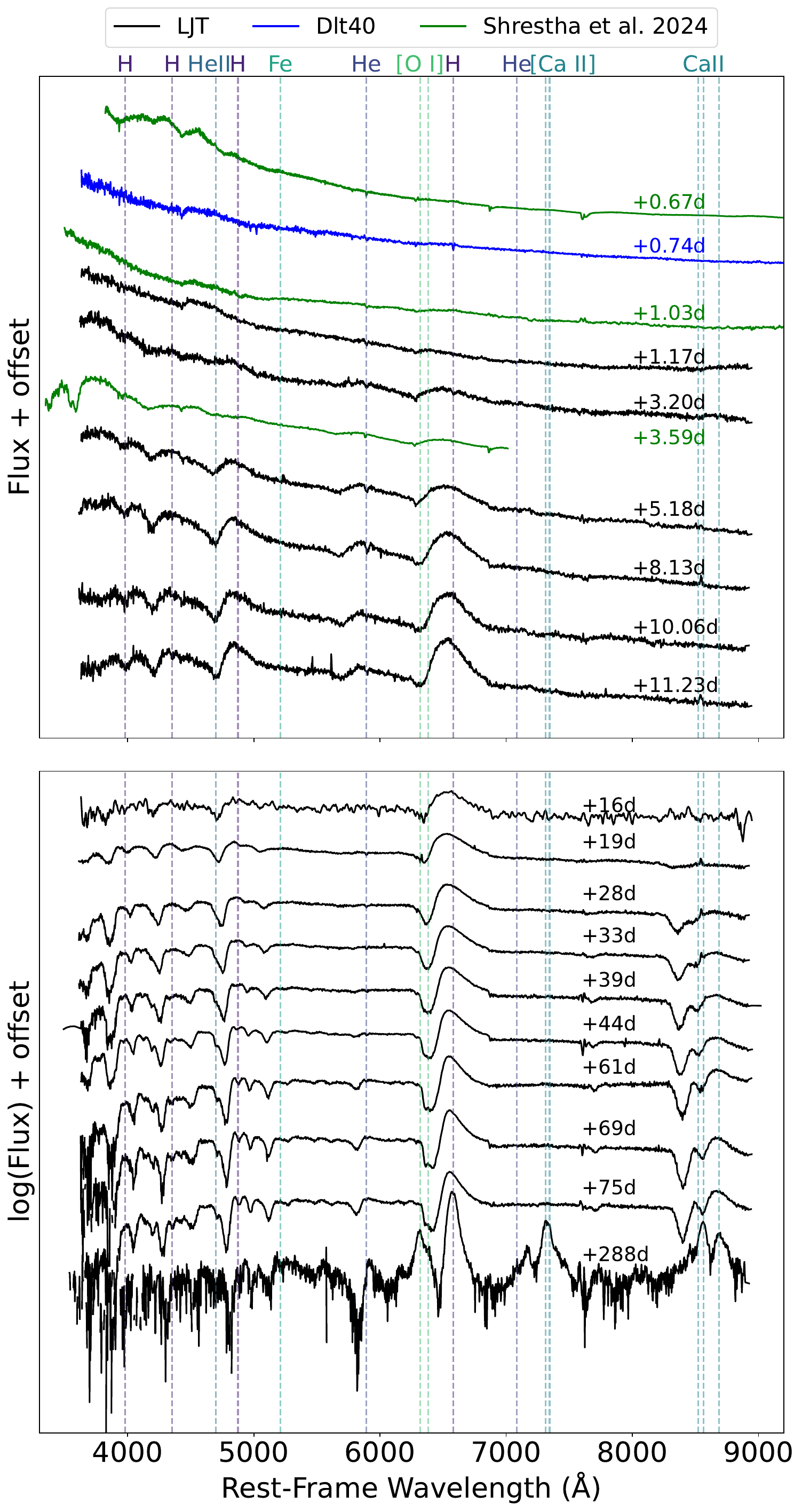}
\caption{Spectra for SN 2023axu from +0.67 to +288 days after the explosion. All spectra have been vertically offset for clarity, with the epochs relative to the explosion date indicated on the right. Dashed lines in various colors mark the rest wavelengths of prominent spectral features for different elements. The spectra from LJT are marked in black, those from DLT40 in blue, and those from \cite{Shrestha_2024} in green.}
\label{fig:specdata}
\end{figure}

\subsection{Spectroscopy}
A total of 16 spectra of SN 2023axu were obtained with YFOSC on the LJT, covering the period from approximately 1 to 288 d after explosion. Wavelength and flux calibration were performed using standard IRAF procedures. Telluric absorption features were removed by comparison with standard star observations, and atmospheric extinction corrections were applied using local extinction data. To achieve complete phase coverage, the spectral series shown in Figure~\ref{fig:specdata} comprises data from the LJT, supplemented by one spectrum retrieved from the Transient Name Server (obtained by the DLT40 team with the KAST spectrograph on the Lick telescope) and three selected spectra from \citealp{Shrestha_2024}. The inclusion of these external spectra is particularly valuable for analyzing the rapid spectral evolution during the earliest phases. For a number of spectra showing significant flux calibration discrepancies, adjustments were made by referencing multi-band photometric data. The first four spectra of SN 2023axu, obtained within 1.5 d after the explosion, exhibit a distinct ``ledge-like'' feature near 4600 \AA. This feature had disappeared in the spectrum taken at $t\approx 3.2$ d. Clear P-Cygni profiles subsequently developed, becoming evident approximately one week post-explosion.

\section{ANALYSIS AND RESULTS}
\label{sec:analysis}

\begin{figure*}[ht!]
\plotone{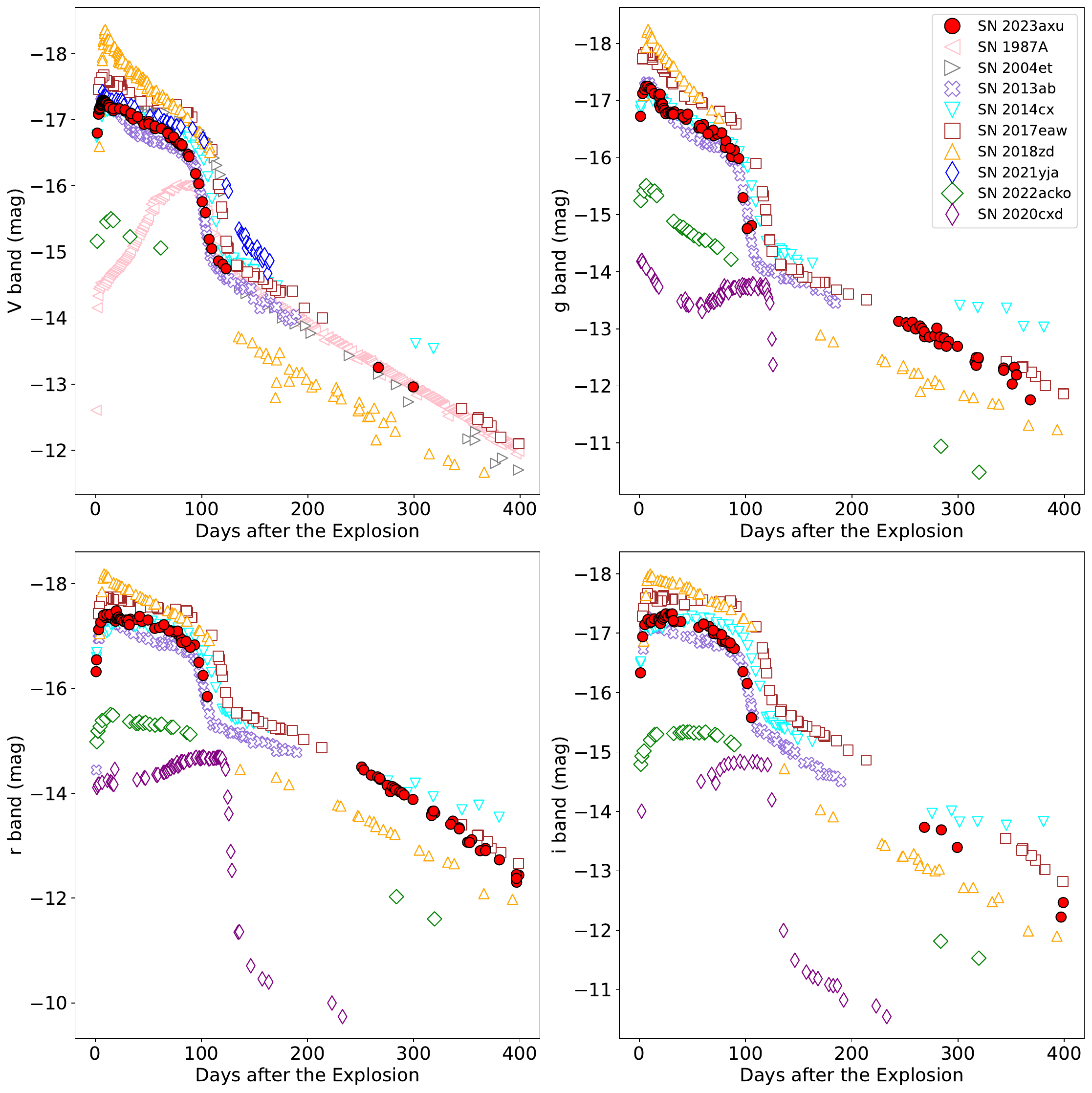}
\caption{Absolute V, g , r, and i-band light curves of SN 2023axu compared with some well-studied SNe II.} 
\label{fig:lightcompare}
\end{figure*}

\subsection{Photometric Analysis}\label{sec:Photometric_Analysis}

The early light curve of SN 2023axu shows a sharp rise in all optical bands. For example, it reaches the V-band peak at about 7 d after the explosion. Adopting the assumed distance to be 13.68 $\pm$ 2.05 Mpc and the total extinction $E(B-V)_{\rm total} = 0.398\pm 0.015$ mag derived by \cite{Shrestha_2024}, the V-band peak of SN~2023axu is $M_{\rm V}= -17.25\pm0.06$ mag. Subsequently, the $V,r,i$ and $z$- band light curve has entered the plateau phase, sustained by hydrogen recombination, right after the peak at about 7 to 10 d after the explosion. Conversely, the g and B-bands exhibit a nearly linear decline rather than a clear plateau. At approximately 90 d after the explosion, the light curves across all bands entered a phase of rapid decline, signaling the end of the plateau phase and the beginning of the tail phase, which is primarily powered by the radioactive decay of synthesized $^{56}$Ni.


Figure~\ref{fig:lightcompare} compares the light curve of SN 2023axu with those of several well-studied SNe II, including SNe~2017eaw \citep{Szalai_2019}, 2018zd \citep{10.1093/mnras/staa2273}, 2020cxd \citep{2022MNRAS.513.4983V, refId0}, 2004et \citep{10.1111/j.1365-2966.2010.16332.x}, 2022acko \citep{10.1093/mnras/staf893}, 2021yja \citep{Vasylyev_2022}, 2014cx \citep{Huang_2016}, and 2013ab \citep{10.1093/mnras/stv759}. The light-curve evolution of SN 2023axu closely resembles that of typical SNe IIP, such as SNe~2017eaw, 2021yja, 2013ab, and 2014cx. This resemblance suggests that the energy source for SN 2023axu is similar to that of the these events. During the nebular phase, the luminosity of SN~2023axu is lower than that of SN 2014cx and closer to that of SN 2017eaw in the $i$- and $g$-bands. However, in the r-band, the luminosities of these three SNe are similar.

\begin{figure*}[]
\includegraphics[width=17.5cm,angle=0]{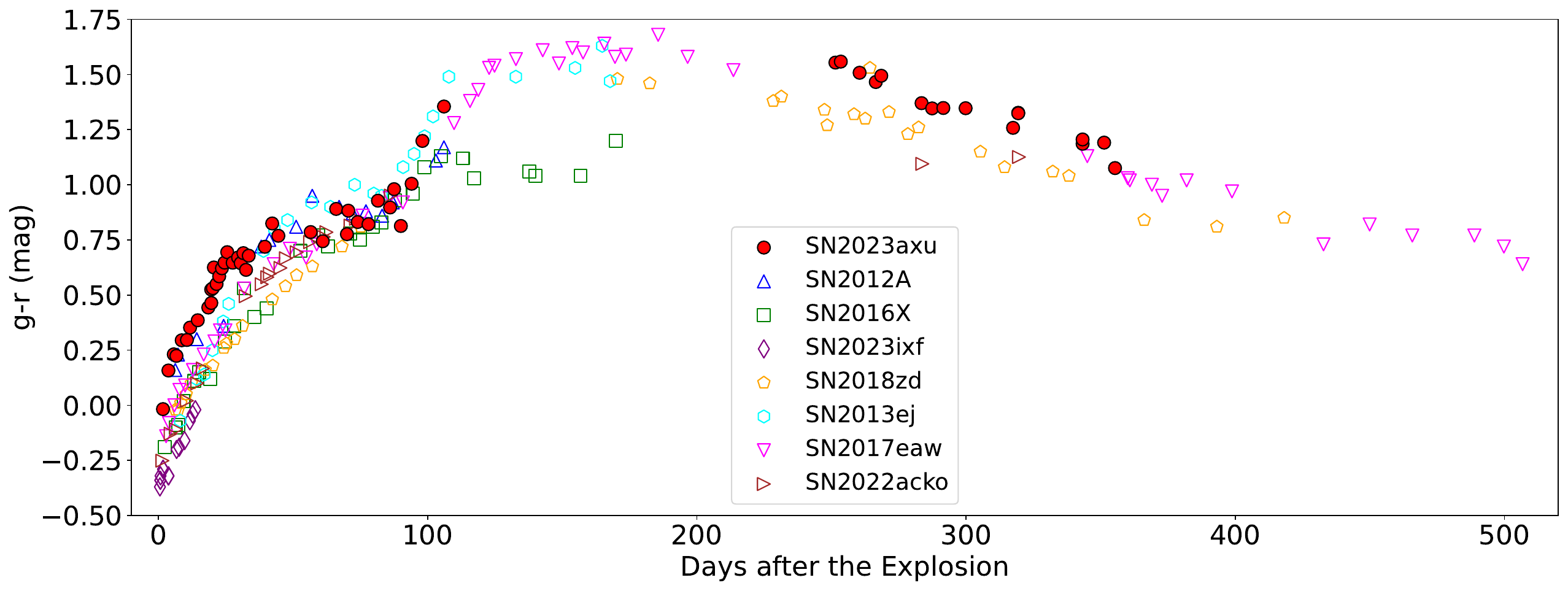}
\caption{The g-r color curve evolution of SN 2023axu along with that of other well-studied.}
\label{fig:g-rcolor}
\end{figure*}

In Figure~\ref{fig:g-rcolor}, the $g-r$ color of SN 2023axu was compared with that of a sample of SNe II. We found that in the early stages post-explosion, SN 2023axu exhibited bluer colors than the comparison sample, resembling SN~2012A and SN 2017eaw. However, after approximately 60 days, the rate of color evolution slowed, causing the g–r color of SN 2023axu to converge towards that of the comparison SNe. In the late stages of the explosion, we observed that SN 2023axu became significantly redder, with a color decline steeper than that of SN 2018zd and comparable to SN 2017eaw. Overall, the color evolution of SN 2023axu is similar to that of SN 2017eaw, which also displays ledge feature in it's early time spectra.

\begin{figure}[ht!]
\includegraphics[width=8.4cm,angle=0]{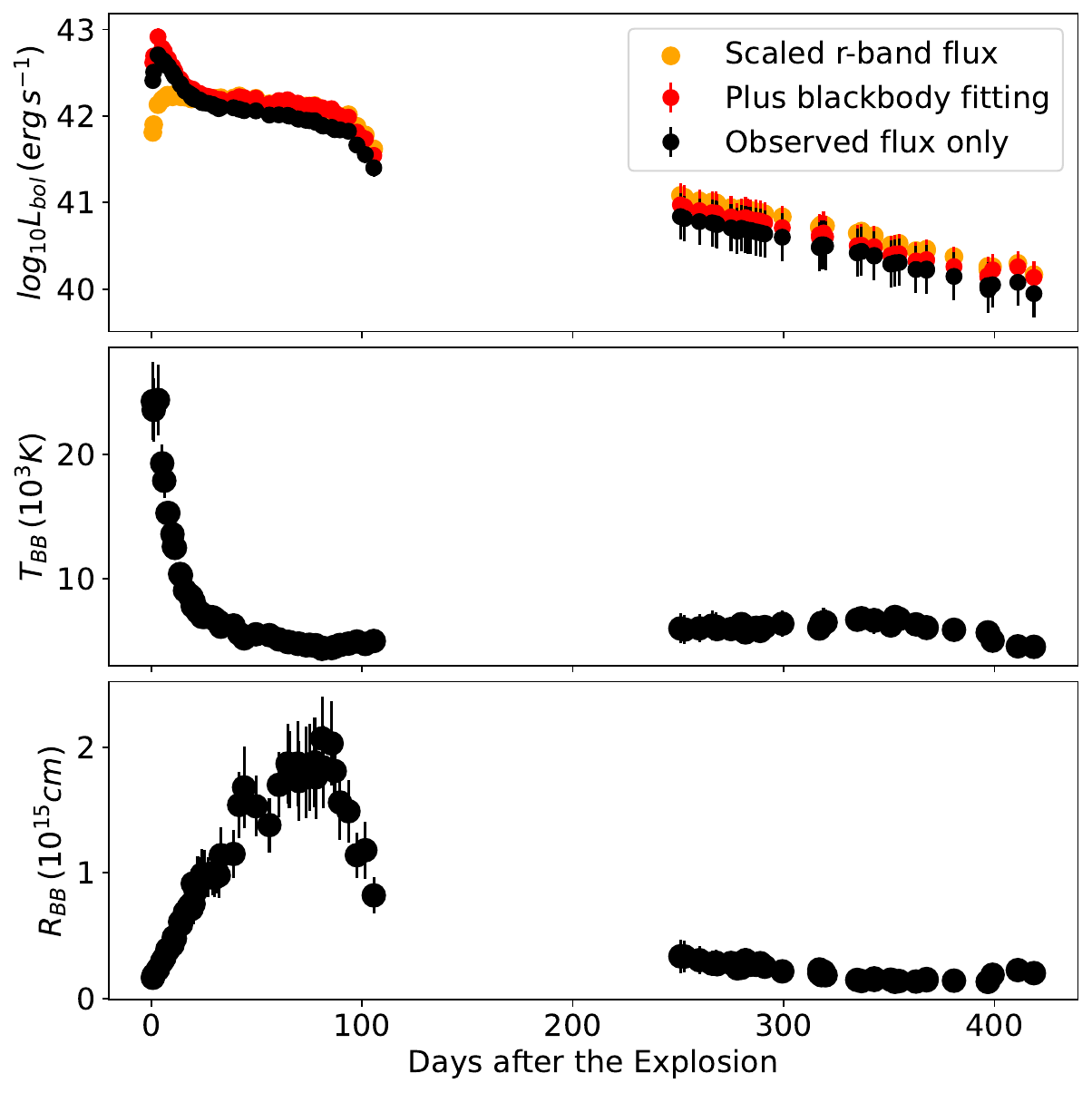}
\caption{Bolometric light curve of SN 2023axu, including the pseudo-bolometric light curve (obtained by integrating the flux in the observed bands), the bolometric light curve derived from blackbody fitting and the r-band flux (scaled by its plateau flux match the calculated plateau flux of the pseudo-bolometric luminosity) as a reference. The middle panel presents the photospheric temperature from the blackbody fit, and the bottom panel shows the resulting photospheric radius, calculated from the bolometric luminosity and temperature. All quantities were derived using the \texttt{Superbol} code.}
\label{fig:bolo}
\end{figure}

\subsection{Bolometric Luminosity and Nickel Mass }
Figure \ref{fig:bolo} displays the ultraviolet-optical-near-infrared pseudo-bolometric light curve of SN 2023axu, constructed by integrating the flux from the multi-band light curves presented in this work, along with the UV light curves obtained from \textit{Swift} observations \citep{Shrestha_2024}, using the SuperBol program\footnote{\url{https://github.com/mnicholl/superbol}} \citep{2018RNAAS...2..230N}. Additionally, the bolometric light curve is derived from these observations under the blackbody assumption. The bolometric luminosity peaks at $L_{\rm max}= (3.8\pm0.3)\times 10^{42}$\,erg\,s$^{-1}$ on MJD = 59978.3, approximately 7 d after the explosion, consistent with the optical light curve.

\begin{figure*}[ht!]
\plotone{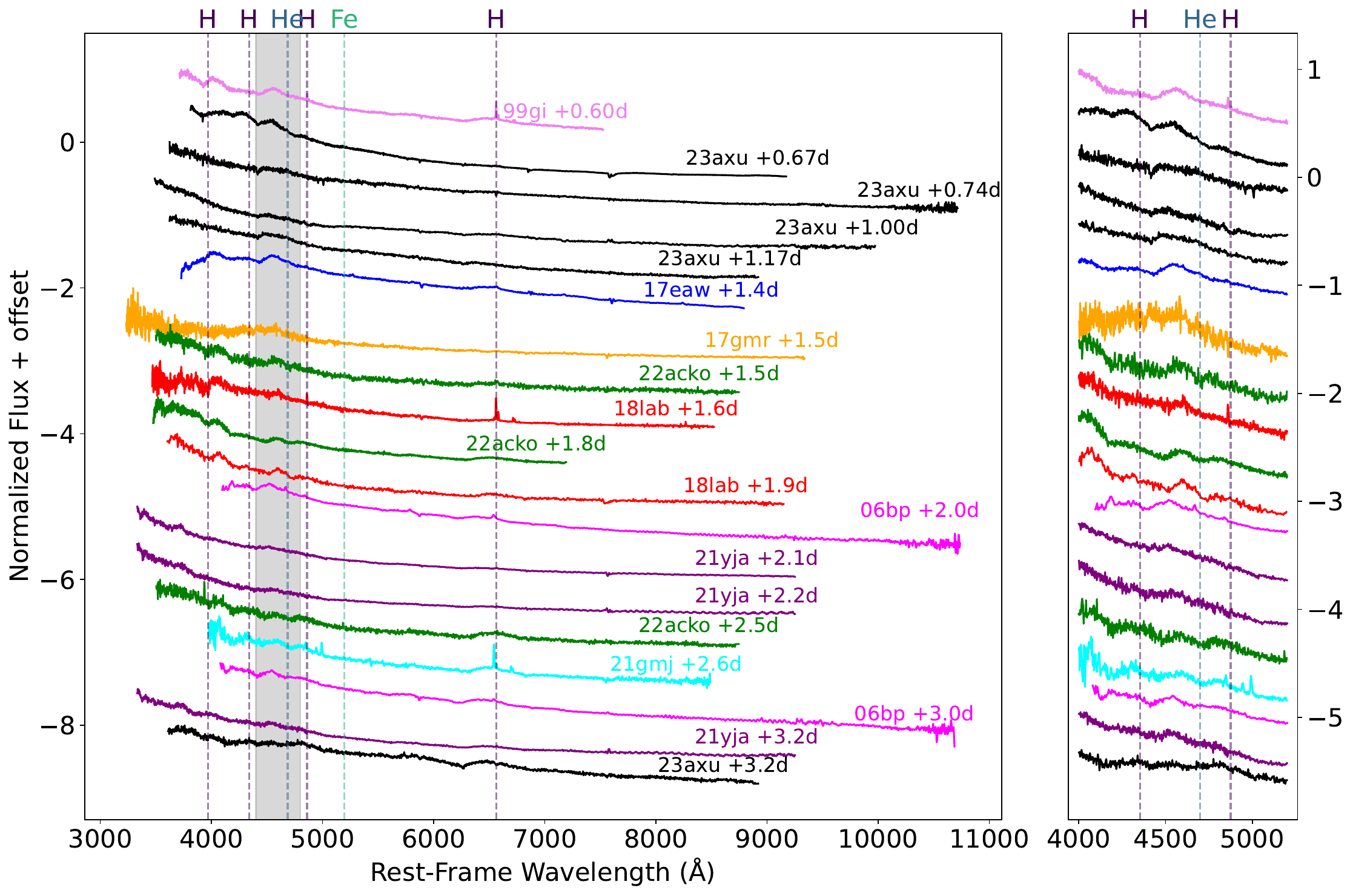}
\caption{Left panel: Comparison of the early spectra of SN 2023axu with other well-studied SNe II exhibiting a similar ledge feature (marked in the gray shaded area in the left panel). Right panel: ledge feature present around 4600 \AA.}
\label{fig:spec_compare}
\end{figure*}

The radioactive-decay tail in SNe II is powered by the decay chain $^{56}$Ni$\to$ $^{56}$Co$\to$ $^{56}$Fe. Thus, a reliable indicator of the $^{56}$Ni mass produced in the explosion is the pseudo-bolometric luminosity. Equation (2) in \cite{2003ApJ...582..905H} provides a method for calculating the mass of $^{56}$Ni by calculating the energy produced from the decay of $^{56}$Ni:
\begin{equation}
M_{Ni} = 7.866 \times10^{-44}\times L_t \times exp{\frac{(t_t-t_0)/(1+z)-6.1}{111.26}M_\odot}
\end{equation}
where $t$ is the phase after explosion in days, $L_t$ is the tail-phase luminosity in units of ergs$^{-1}$, and z is the redshift of the SN. We applied this method to estimate the mass of $^{56}$Ni produced by the explosion to be $M_{Ni} = 0.058^{+0.017}_{-0.013} M_\odot$, in excellent agreement with the value of 0.058\Msun\, obtained via a similar method by \citet{Shrestha_2024}. This estimate is comparable to the nickel mass of $0.056 \pm 0.008 M_\odot$ reported for SN 2014cx \citep{Huang_2016}, consistent with the similar late-time light-curve evolution noted in Section~\ref{sec:Photometric_Analysis}.

For the effective calculation, we assume that the energy released by radioactive decay is fully captured, the deposited energy is re-emitted instantly, and no other energy sources are present. However, we calculate that the pseudo-bolometric decline rate during this period is 1.08 mag 100 day$^{-1}$ , which is higher than the complete capture rate of $^{56}$Co decay, estimated at 0.98 mag 100 day$^{-1}$. This indicates a loss from the energy generated by radioactive decay, so the actual mass of $^{56}$Ni produced could be higher than our calculated value.

We can also estimate the $^{56}$Ni mass from the nebula spectrum. \cite{2003MNRAS.338..939E} noted a correlation between H$\alpha$ intensity and nickel mass, and \citet{2012MNRAS.420.3451M} provided the relation:
\begin{equation}
  M_{\rm Ni} = A \times 10^{B \times {\rm FWHM}_{\rm corr}}\,
  {\rm M}_{\odot}\, ,
\end{equation}
\noindent
where $A=1.8_{-0.7}^{+1.0} \times 10^{-3}$ and $B=0.023\pm0.004$, FWHM$_{\rm corr}$ is the full width at half-maximum intensity corrected for the instrumental broadening effect. Fitting the H$\alpha$ of nebular spectrum of SN 2023axu ($t \approx$ 288 d) gives FWHM$_{\rm corr}\approx63.2$ \AA, yielding $M_{Ni} = 0.05^{+0.02}_{-0.01}\ \rm M_\odot$, consistent with the value derived from late-time luminosity. By averaging the values derived from spectroscopic and photometric methods, we obtain a nickel mass estimate of $M_{Ni} \approx 0.055\ \rm M_\odot$.

\subsection{Spectral Analysis}
\label{subsec:SpectralAnalysis}
The most distinctive feature in the early-time spectra of SN 2023axu is a ledge-like, broad emission peak near 4600 \AA, which has been observed in only a handful of SNe II \citep{Andrews_2019,10.1093/mnras/staf893,Rehemtulla_2025,Pearson_2023,10.1093/mnras/stae170}. Figure \ref{fig:spec_compare} compares the early spectra of SN 2023axu (at $t < 4$ d) with those of other well-studied SNe II that exhibit ledge feature at similar phases, including SNe 2017eaw \citep{Szalai_2019}, 2021yja \citep{Hosseinzadeh_2022}, 1999gi \citep{Leonard_2002}, 2018lab \citep{Pearson_2023}, 2017gmr \citep{Andrews_2019}, 2022acko \citep{10.1093/mnras/staf893}, 2006bp \citep{Quimby_2007}, and 2021gmj \citep{10.1093/mnras/stae170}. The ledge feature of SN~2023axu resembles those seen in SNe 2017eaw, 2006bp, and 2021yja.

The early spectra of SNe 2022acko and 2006bp showcase a narrow H$\alpha$ emission line superimposed on a broad component. In contrast, no such feature is present in the early spectrum of SN 2023axu. A very faint H${\alpha}$ P-Cygni emission appeared in the spectrum 1.17 d post-explosion, with a velocity reaching 15,000 km s$^{-1}$ derived from the absorption minimum. The ledge feature becomes faint in the spectrum taken 3.2 d post-explosion, while the hydrogen emission lines become clearer, maintaining a velocity around 15,000 km s$^{-1}$. Additionally, the H$\beta$ and H$\gamma$ lines are observed, also at velocities close to 15,000 km s$^{-1}$. If the ledge is considered to be the He II $\lambda$4686 emission line \citep{Bullivant_2018, Andrews_2019}, its velocity is approximately 20,000 km s$^{-1}$. A detailed discussion of the ledge feature in SN~2023axu is provided in Section \ref{subsec:ledge}.

Following the disappearance of the ledge feature, the spectra of SN 2023axu evolved into those typical of SNe~IIP, characterized by P-Cygni profiles of hydrogen and helium superimposed on a cooling blackbody continuum. At $t \sim$33 d, the H$\alpha$ absorption trough had expanded and developed a square-shaped profile. Concurrently, the overall P-Cygni profile widened, exhibiting a shallow absorption component at approximately 10,000 km s$^{-1}$ of unknown origin. This feature, termed the ``Cachito" \citep{Gutierrez_2017}, is located within the broad absorption structure. Subsequently, at $t\sim$ 40~d, metal lines such as Ba II and Fe II, along with various blends, became distinct in the spectra.
\begin{figure}[ht!]
\includegraphics[width=8.4cm,angle=0]{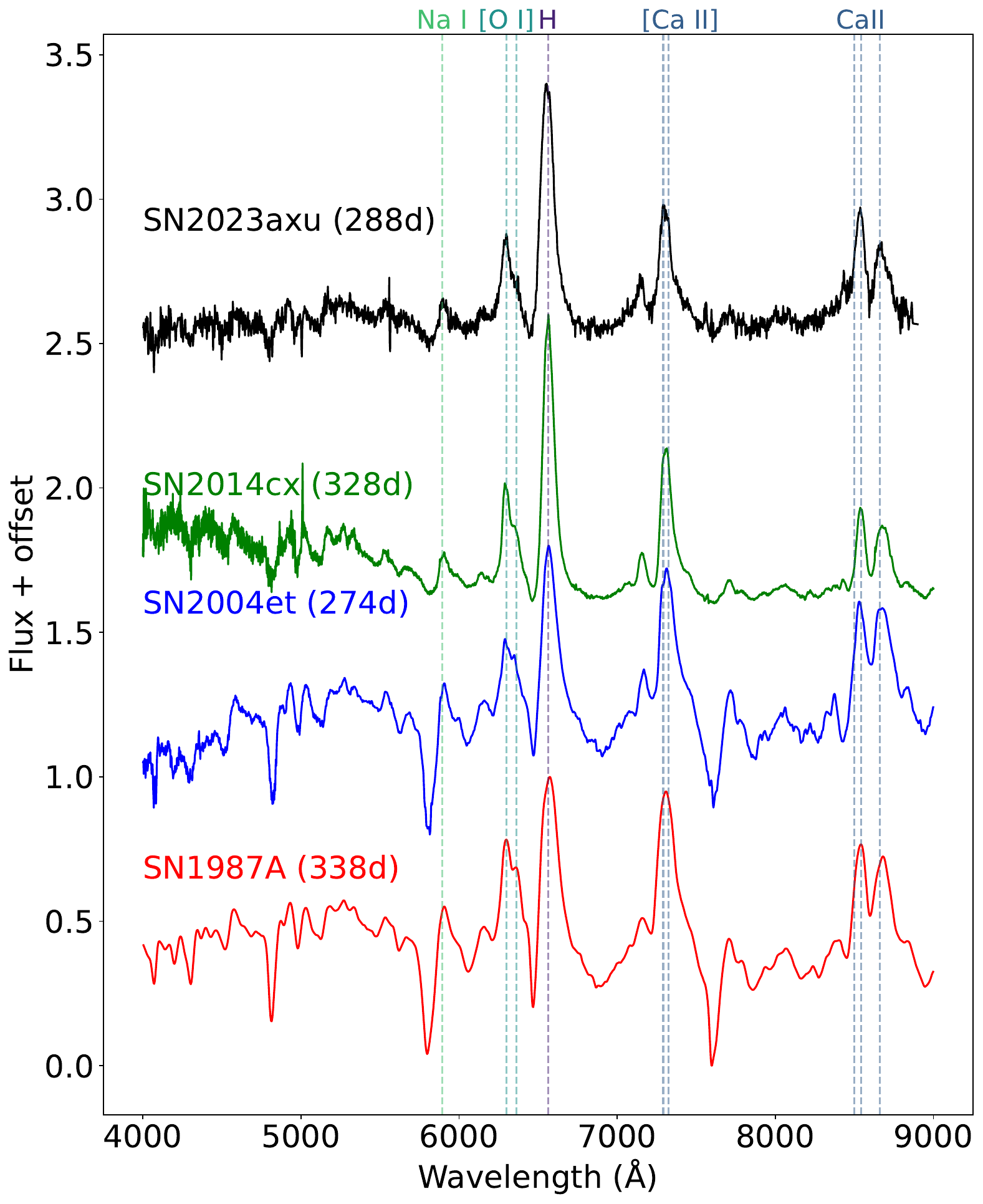}
\caption{Comparison of the reddening- and Doppler-corrected spectra of SNe 2023axu, 2014cx, 2004et and 1987A obtained at about 300 d after the explosion.}
\label{fig:latespec}
\end{figure}

In Figure \ref{fig:latespec}, we compare the late time spectra of SN 2023axu with SNe 2014cx \citep{Huang_2016}, 2004et \citep{10.1093/mnras/stu955} and 1987A \citep{pun1995ultraviolet}. We found that the late spectral features of SN 2023axu are more similar to those of SN~2014cx, while being weaker than those of SN 2004et and SN 1987A. The most noticeable difference is that the Na~I P-Cygni profiles of SN 2023axu and SN 2014cx exhibit shallower absorption compared to the strong absorption seen in SN 2004et and SN 1987A. Additionally, the Ca~II NIR triplet line emission is stronger in the latter two SNe (SN 2004et and SN 1987A), potentially indicating a higher degree of ionization.


\subsection{Late Time Evolution}\label{sec:lateevolution}

\begin{figure}[]
\includegraphics[width=8.4cm,angle=0]{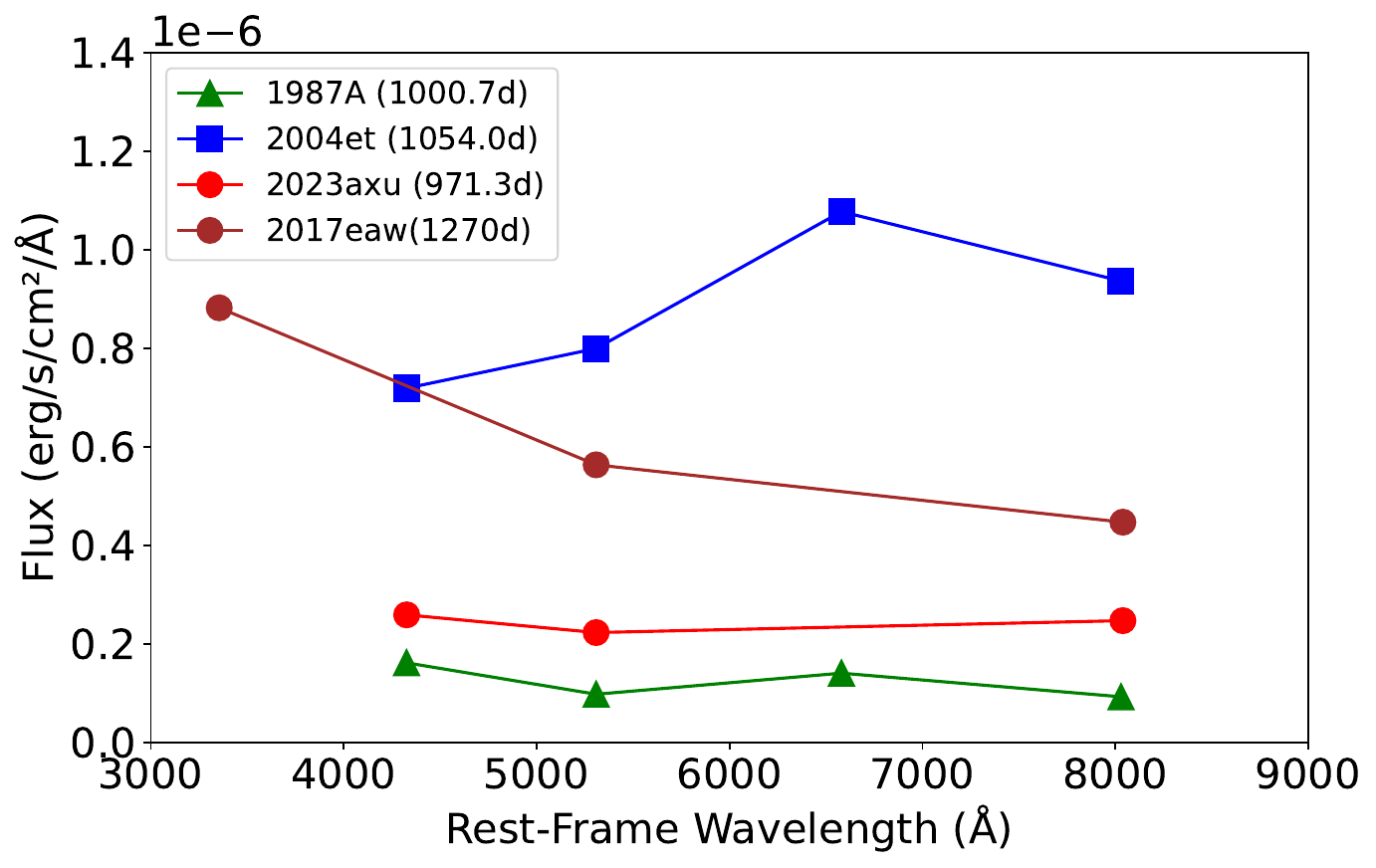}
\caption{Late time SED of SN 2023axu with those of SN 1987A, SN 2004et, and SN 2017eaw at approximately 1000 d post-explosion. The SED of SN 2023axu is derived from HST photometry in the F438W, F555W, and F814W filters, In comparison, the data for SN 2017eaw is derived from HST photometry in the F336W, F555W, and F814W filters. Meanwhile, the photometric data for SN 2004et and SN 1987A are obtained in Johnson $BVRI$- bands. The data of SN 2017eaw is from \cite{10.1093/mnras/stac3549},  the data of SN 2004et is from \cite{10.1111/j.1365-2966.2010.16332.x}, and the data of SN 1987A is from the Open Supernova Catalog\footnote{https://github.com/astrocatalogs/supernovae} \citep{Guillochon_2017}.}
\label{fig:SED}
\end{figure}

\begin{figure}[]
\includegraphics[width=8.4cm,angle=0]{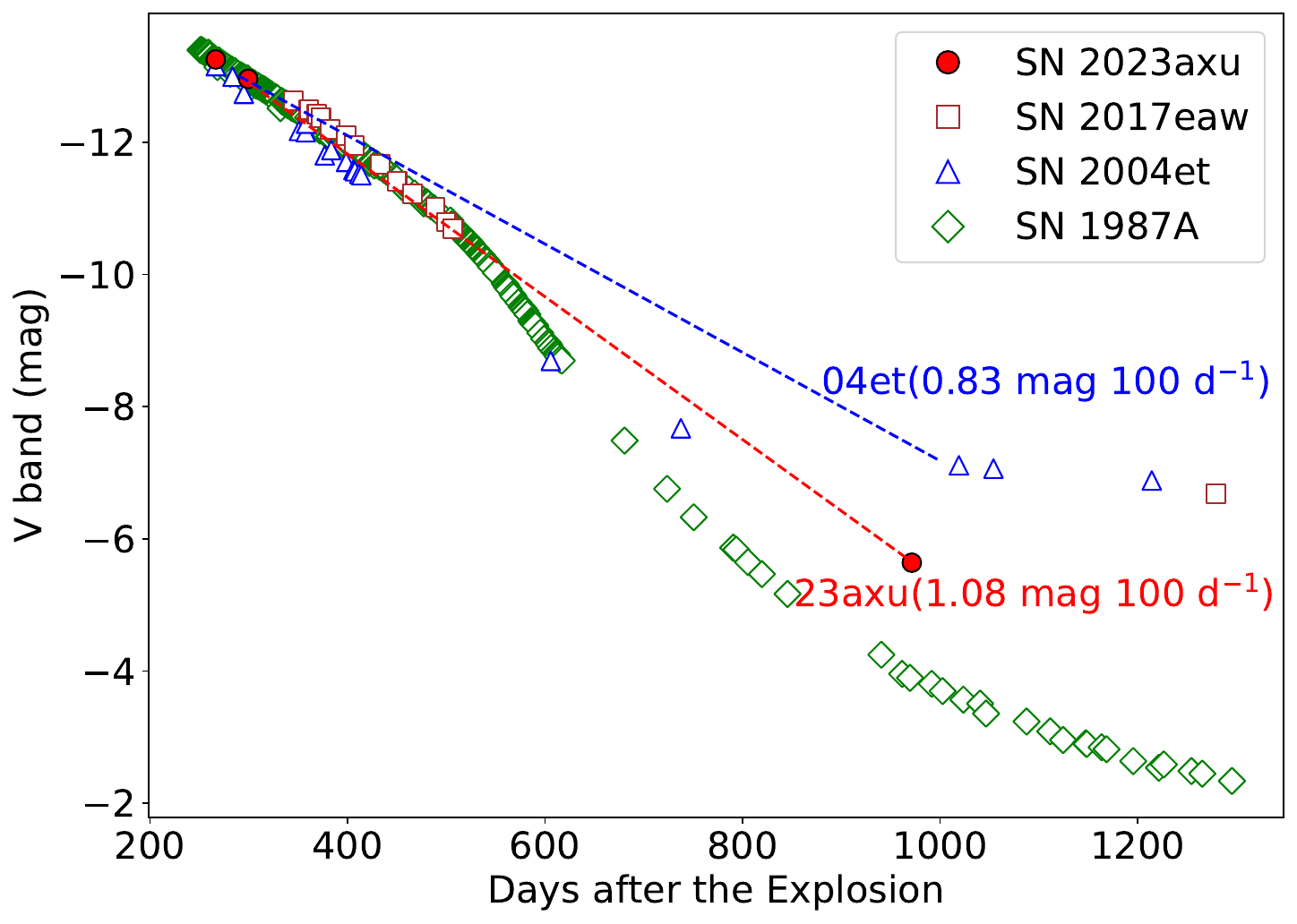}
\caption{Comparison of the late-time absolute V-band light-curve evolution of SN 2023axu with SNe 2017eaw, 2004et, and 1987A. For reference, the decline rate of SN 2023axu and SN 2004et between  300 and 1000 d post-explosion is indicated with dashed lines.}
\label{fig:V_band_compare_lines}
\end{figure}

To gain deeper insights into the late-phase luminosity evolution of SN 2023axu, we accessed HST observations data at approximately 1000 d post-explosion from the MAST. From these data, we derived absolute fluxes to construct its spectral energy distribution (SED) at this epoch. Figure~\ref{fig:SED} presents this SED alongside contemporaneous SEDs of SN 1987A, SN 2004et and SN 2017eaw, showing that the SED of SN 2023axu closely resembles that of SN 1987A, whereas SN 2004et and SN 2017eaw exhibit brighter SED.

The F555W band is photometrically equivalent to the Johnson V-band, so we treated the F555W magnitudes as V-band measurements to compute the decline rate between 300 and 1000 d post-explosion. Over this interval, the late-time luminosity is dominated by the radioactive decay of $^{56}$Co. We derive a decline rate of 1.08 mag 100 d$^{-1}$ for SN 2023axu, significantly steeper than those of SN 2004et (0.83 mag 100 d$^{-1}$) and SN 2017eaw (0.64 mag 100 d$^{-1}$). Note that the rate for SN 2017eaw is measured over a longer baseline (300–1270 d), which may introduce additional uncertainty.
Moreover, observational evidence indicates that the ejecta of SN 2017eaw interacted with pre-existing CSM at $\sim$ 900 d \citep{Weil_2020}, which would flatten the late-time decline and thus affect the comparison.

However, a comparison the $V$- band light curves in Figure \ref{fig:lightcompare} shows that up to $t\approx 400$d the luminosity evolution of SN 2023axu closely tracks that of SN 2017eaw and SN 2004et, with an overall late-time decline rate near the $^{56}$Co decay rate (0.98 mag 100 d$^{-1}$). This suggests that the differences in decline rates  among these SNe emerges only after $\sim$400 d,  i.e., within the latter portion of the 300–1000 day interval.

A plausible explanation for this divergence is sustained interaction with a circumstellar stellar wind.  \cite{dessart2023morphing} presents radiative-transfer calculations for a SN II with and without such interaction, focusing on 350–1000 d after the explosion.  In models without interaction, the ejecta are powered solely by radioactive decay, producing an exponential luminosity decline. In contrast, with continuous interaction against a steady wind, the contribution from radioactive decay diminishes beyond $\sim$ 350 d while the thermalized shock power from ejecta–CSM interaction grows, becoming dominant by $\sim$ 1000 d. This leads to a progressive flattening of the very-late-time light curve, as seen in SN 2017eaw and SN 2004et. The absence of comparable flattening in SN 2023axu therefore suggests minimal wind interaction at late times, making it a compelling example of a SN II with negligible late-time interaction power.

\section{DISCUSSION}
\label{sec:discussion}

\subsection{Origin of Ledge Feature}
\label{subsec:ledge}

\begin{figure*}[ht!]
\includegraphics[width=17.4cm,angle=0]{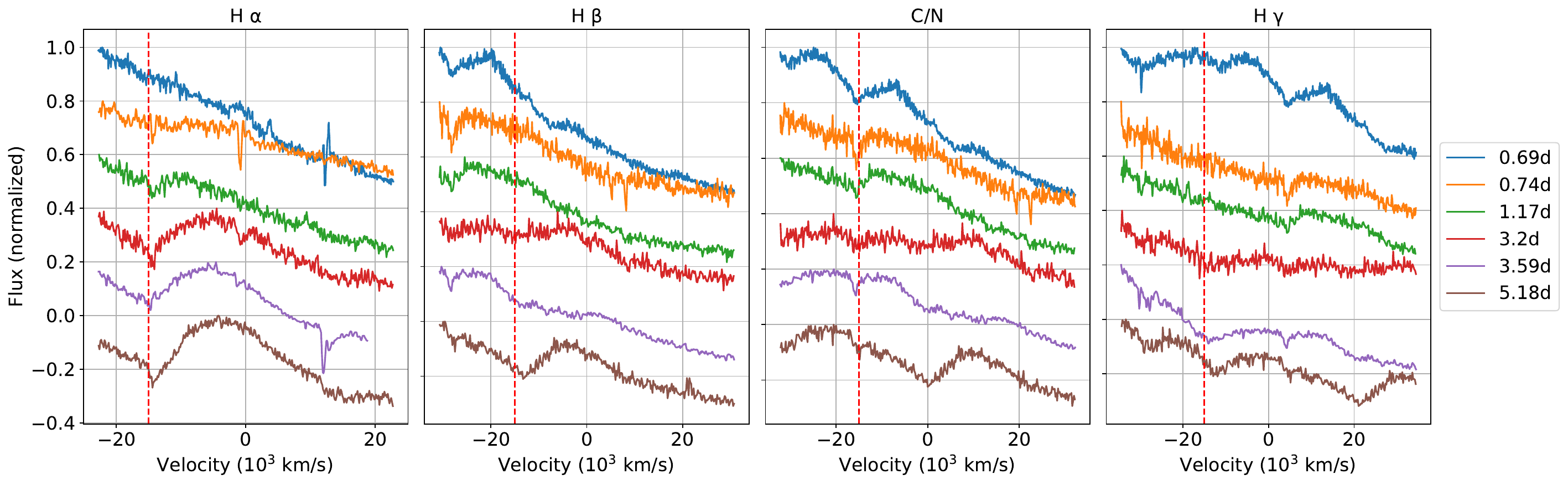}
\caption{The measurement of emission line velocities in the early spectra of SN 2023axu included the Balmer emission lines of H and the group of C III and N III emission lines (with $\lambda$4645 Å selected for measurement), all spectra were normalized. The red dashed line indicating a velocity of 15,000 km s$^{-1}$ as a reference. There is an atmospheric absorption line near the H$\alpha$ line at 6280 \AA, which coincides closely with the position of its minimum absorption. Although there is still a slight effect after atmospheric extinction correction, it does not affect the judgment of its velocity.}
\label{fig:earlyv}
\end{figure*}

Explanations for the origin of the ledge feature are diverse but can be broadly classified into three categories. \cite{10.1111/j.1365-2966.2006.10587.x} attributes this structure to high-velocity (HV) H$\beta$. However, our measurements of the Balmer line velocities (Figure \ref{fig:earlyv}) indicate a maximum hydrogen velocity of approximately 15,000 km s$^{-1}$. 
At 1 d after the explosion, when the ledge feature is clearly visible, is difficult to precisely determine the velocities of the H$\beta$ and H$\gamma$ lines. Nevertheless, based on the velocity of the H$\alpha$ line, they are unlikely to exceed 20,000 km s$^{-1}$. If the ledge feature were due to HV H$\beta$, its velocity would need to be as high as 30,000 km~s$^{-1}$, which significantly exceeds our observed values. We therefore do not favor this interpretation.

\cite{Quimby_2007} suggest that this feature may instead arise from He II $\lambda$4686, which would require high ionization in the outer ejecta layers. This interpretation has also been applied to SNe 2010id \citep{Gal-Yam_2011}, 2017gmr \citep{Andrews_2019} and 2013fs \citep{Bullivant_2018}. For example, the early spectrum of SN 2006bp exhibited narrow He II $\lambda$4686 emission with electron-scattering wings. Similar narrow emission features have been reported in other SNe such as SN 2013fs \citep{Bullivant_2018}. However, no such feature is present in the early spectra of SN 2023axu.


A third explanation posits that the ledge consists of broad lines produced by bulk motion, blended with several ionized features from CSM \citep{Hosseinzadeh_2018,Hosseinzadeh_2022,Soumagnac_2020,Bruch_2021}. The feature may be a blend of ions such as N V, N III, C III, O III, and He II; approximate velocities for these lines are indicated in Figure \ref{fig:earlyv}. Narrow emission lines with electron-scattering wings result from non-coherent scattering by thermal electrons, whereas bulk motion can produce broad, blended spectral features \citep{10.1111/j.1365-2966.2008.14042.x}. However, as noted by \citet{Hosseinzadeh_2022}, the breadth and strength of the ledge may be inconsistent with an origin in multiple narrow emission lines.

\begin{figure}[ht!]
\includegraphics[width=8.4cm,angle=0]{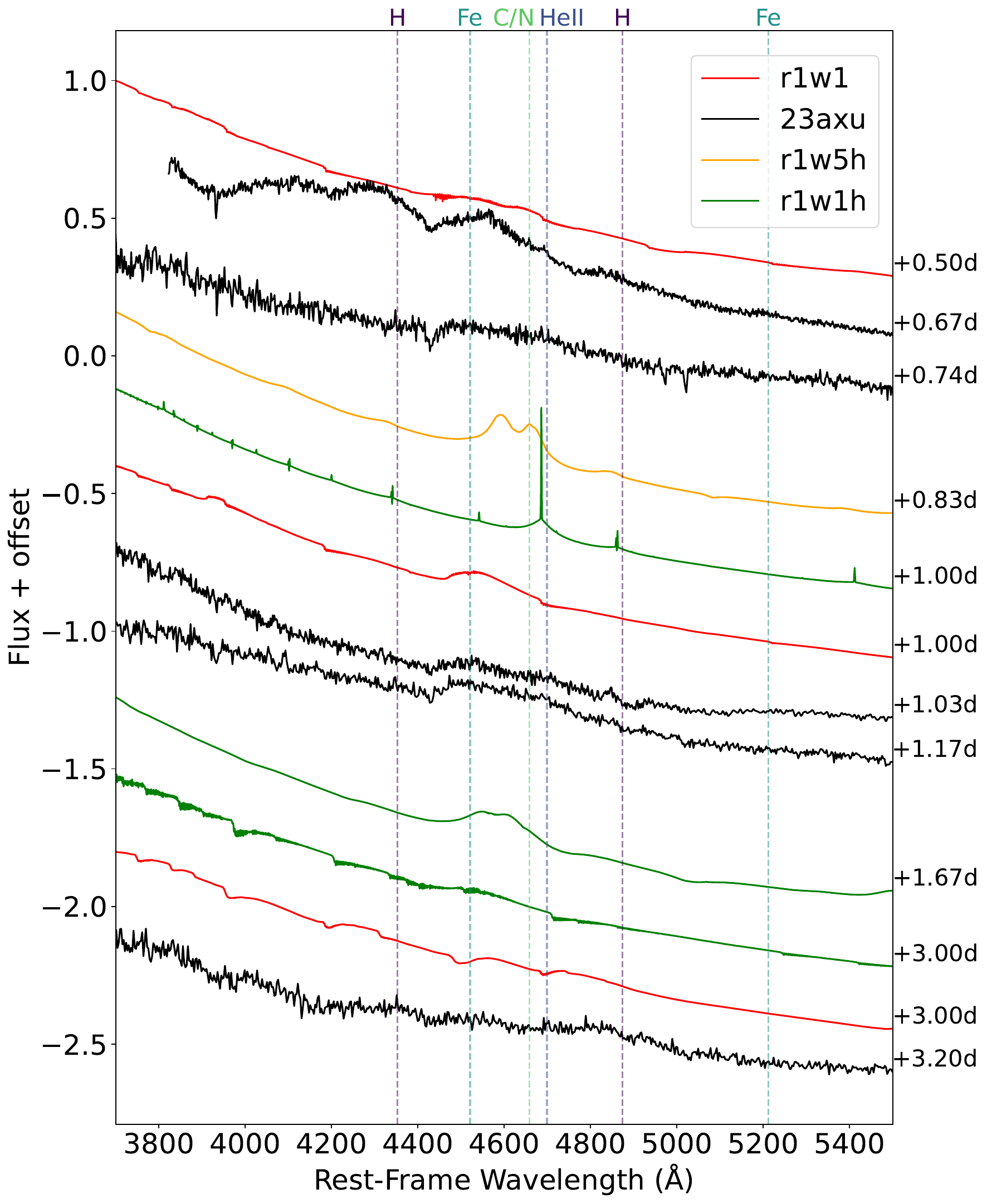}
\caption{Comparison of the early spectra of SN~2023axu with the r1w1, r1w1h, r1w5h model spectrum from \cite{DessartLuc}. The r1w1 and r1w1h model have a wind mass loss rates of $10^{-6}M_{\odot}\,yr^{-1}$ while r1w5h model has a wind mass loss rates of $5\times10^{-3}M_{\odot}\, yr^{-1}$, The r1w1h and model include an extended RSG atmosphere of $H_{\rho}=0.3\, R_{\star}=150.3\, R_{\odot}$, and the r1w5h model's atmosphere is $H_{\rho}=0.1\, R_{\star}=50.1\, R_{\odot}$, where $R_{\star}$ is the radius of RSG. }
\label{fig:luc}
\end{figure}

We also compare SN 2023axu with the synthetic spectra from the multi-group radiation-hydrodynamics and non-LTE radiative transfer models of \citet{DessartLuc}. These models depend on three key parameters: the initial RSG radius (r), the wind mass-loss rate (w), and the presence of an extended atmospheric envelope (h). \citet{Pearson_2023} and \citet{Hosseinzadeh_2022} matched observed spectra with models r1w1h and r1w5h, suggesting that an RSG exploding into a low-density CSM with an extended atmosphere can account for the early spectra of SN 2018lab and SN 2021yja. In Figure \ref{fig:luc}, we compare the early spectra of SN 2023axu with the \citet{DessartLuc} models (r1w1, r1w1h, and r1w5h). The spectrum at 0.67 d is not well reproduced by any model, as it shows additional broad features between 4000 to 4400 \AA\,that are absent from the model grid. By contrast, the spectra at 0.74, 1.03, and 1.17 d most closely resemble the r1w1 model at 1.0~d and the r1w1h model at 1.6 d, consistent with \cite{Shrestha_2024}. However, r1w1h predicts narrow emission lines at early times that are not observed in SN 2023axu. Furthermore, our comparison with the pseudo-bolometric light curve in Section \ref{subsec:Diversity} indicates that r1w1 provides a better match. We therefore favor r1w1 as the best overall fit, although we cannot entirely rule out the presence of weak, undetected narrow features in the earliest spectrum. 

\begin{figure*}[ht!]
\plotone{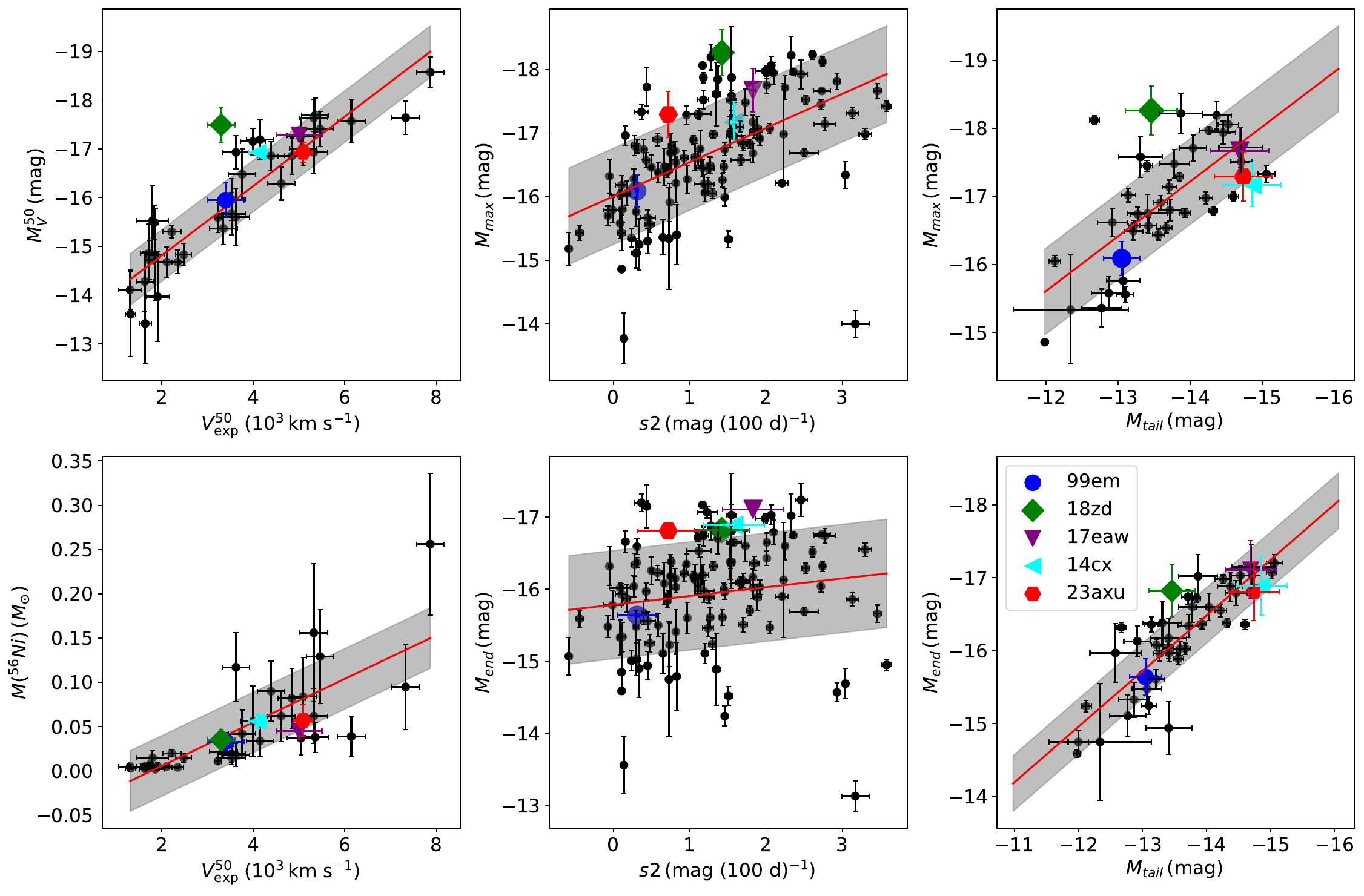}
\caption{Photometric and spectroscopic parameter space for SNe II, showing the position of SN 2023axu relative to a literature sample \citep{2003ApJ...582..905H,Anderson_2014,Spiro_2014,Zhang_2014}. Parameters include the V-band absolute magnitude measured at around 50 d after explosion($M^{50}_{V}$), the expansion velocity of Fe II λ5169 measured at 50 d after the explosion ($v^{50}_{exp}$), the mass of $^{56}$Ni, three shape parameters of the V-band light curve defined by \cite{Anderson_2014}($M_{max}$,$M_{end}$,$M_{tail}$), and the decline rate of the plateau (s2). Individual SNe (1999em, 2018zd, 2017eaw, 2014cx) are highlighted. The red line and gray band show the linear fit and its $1\sigma$ uncertainty.}
\label{fig:Vband}
\end{figure*}

Regarding whether the ledge feature results from a blend of multiple emission lines, \cite{10.1093/mnras/staf893} note that H$\alpha$ should also show a characteristic narrow emission component with electron-scattering wings—a feature absent in SN 2023axu and many other SNe. The shape of ledge in SN 2023axu more closely resembles that of SNe 2017eaw and 2021yja, appearing similar to a P-Cygni profile. Furthermore, at $t\sim 0.67$ d, broad emission features appeared in the 4000-4400 \AA\ range, similar to those seen in later spectra. We conclude that these features originate from the outer ejecta. Under rapid expansion, the optical depth of the ejecta changes quickly. Initially, the outer layers are transparent, allowing high-energy spectral lines to appear briefly; within hours, the inner layers become optically thick, causing these emission features to vanish. This rapid evolution also supports the view that the CSM around SN 2023axu is very tenuous, unable to absorb these photons efficiently. The features disappear in the subsequent spectrum taken at 0.74 d, just hours later.

We therefore support the view of \citet{10.1093/mnras/staf893} and \citet{Dessart_2008} that the ledge of SN 2023axu is a P-Cygni feature originating from ionized ejecta—not from flash-ionized unshocked CSM. \cite{10.1093/mnras/staf893} proposes that the ledge feature is blueshifted He II $\lambda$4686 for SN 2022acko, produced by high-energy photons arising from the CSI of SN ejecta which produced high-ionization species. The spectral lines originating from the ionized SN ejecta will exhibit distinct line shapes and velocities compared to those from the CSM, resulting in such a broad feature. The He II $\lambda$4686 line velocity of 2022acko is close to its hydrogen line, supporting the above conclusion. But as noted in section \ref{subsec:SpectralAnalysis}, the ledge of SN 2023axu is more blueshifted, resulting in a higher He II velocity, while the velocities that match better should be those of C III $\lambda$4647 and N~III $\lambda$4638 lines as seen in Figure \ref{fig:earlyv}. Therefore, the ledge feature of SN~2023axu should be result from the overlapping emission features of C III $\lambda$4647, N III $\lambda$4638, and He~II $\lambda$4686 lines.

We propose the physical mechanism underlying formation of the entire ledge feature: During the progenitor phase, minimal mass loss results in the absence of a dense CSM shell. The sparse ejected material remains confined near the progenitor system. Following the SN explosion, the rapidly expanding ejecta interact with this tenuous material, ionizing and accelerating it to extreme velocities (potentially including the progenitor's outer layers). This interaction generates the observed ledge feature, explaining both its Doppler broadening and high-velocity characteristics. Furthermore, the weak-wind scenario implied by the ledge feature is supported by the late-time photometric decline rate and the absence of clear wind-interaction signatures (Section \ref{sec:lateevolution}), together indicating that the progenitor of SN 2023axu did not sustain a strong wind capable of producing a dense CSM.

\begin{figure}[ht!]
\centering
\includegraphics[width=8.5cm,angle=0]{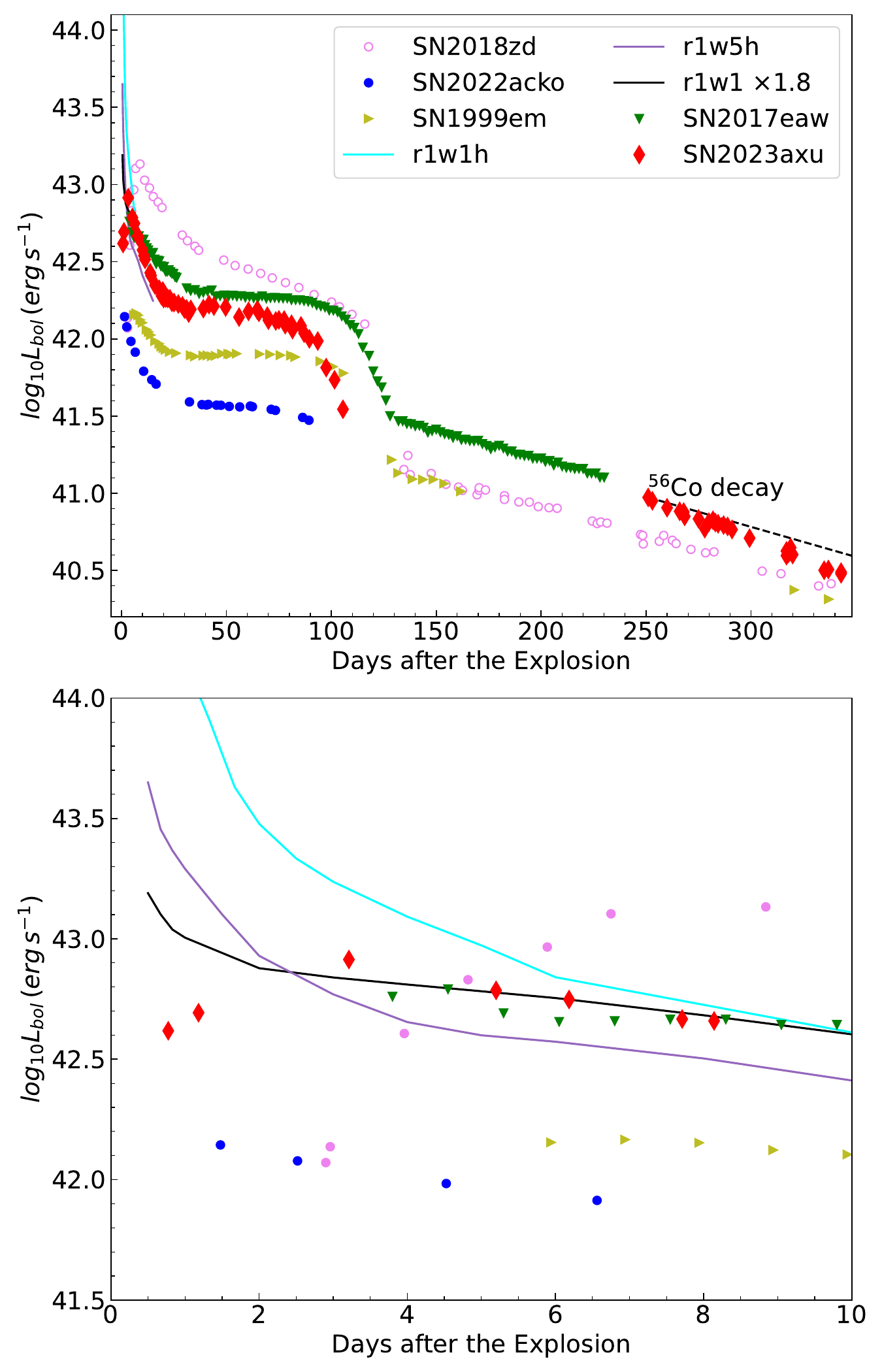}
\caption{Pseudo-bolometric light curve compared with that of a few well-studied SNe II and the model bolometric light curve in the \cite{DessartLuc}.}
\label{fig:bolocompare}
\end{figure}

\subsection{Weak Stellar Wind Interaction of SNe II}
\label{subsec:Diversity}

To contextualize SN 2023axu within the population of SNe II, we compare its spectral and photometric parameters with those of well-studied SNe II from the literature \citep{2003ApJ...582..905H, Anderson_2014, Spiro_2014, Zhang_2014}, as well as with individual objects: the normal SNe II 2014cx and 1999em, the ledge-feature SN 2017eaw, and the flash-ionized SN 2018zd. The comparison, illustrated in Figure \ref{fig:Vband}, reveals the following:

(1) SN 2023axu and SN 2017eaw, which exhibit ledge features, are brighter by $0.75\pm0.5$ mag at peak ($M_{max}$) and in the tail ($M_{tail}$) compared to SNe samples with a similar decline rate of s2 \citep{Anderson_2014}. However, their middle plateau ($M^{50}_{V}$) are consistent with those of SNe having comparable ejecta expansion velocities ($v^{50}_{exp}$). 

(2) The luminosity decline from peak to tail is more gradual in SN 2023axu and SN 2017eaw than in the sample average, being slightly lower and closer to that of typical SNe II, such as SN 2014cx. In contrast, SN~2018zd, which exhibits flash-ionized emission lines, experiences a significant luminosity drop from peak to tail \citep{10.1093/mnras/staa2273}.

A similar trend is seen in the light-curve comparison in Figure \ref{fig:lightcompare}: the ledge-sample objects closely resemble ordinary SNe II, whereas flash-ionized events like SN~2018zd show higher early peak luminosity and a steeper pre-plateau decline. This is expected if flash features arise from CSM interaction, where a substantial fraction of kinetic energy is converted into radiation via strong ejecta–wind collision. The luminosity of SN 2023axu at various phases resembles that of normal SNe II, and its decline rate is relatively slow. This suggests that the efficiency of converting kinetic energy into radiation within the ionized ejecta is low, with only a small fraction of kinetic energy being thermalized, yielding negligible optical brightening.

The light-curve evolution of SN 2023axu stands in sharp contrast to SNe interacting with dense CSM, such as SN 2018zd. The latter typically show narrow emission lines, sustained early brightness, and light-curve shapes resembling SNe IIL. To elucidate the luminosity evolution of SN 2023axu, we compared its pseudo-bolometric light curve with other well-studied SNe II samples and model bolometric light curves from \cite{DessartLuc} in Figure \ref{fig:bolocompare}. While pseudo-bolometric estimates can be affected by band coverage and the non-blackbody nature of SNe, and the \cite{DessartLuc} models are idealized and may not map one-to-one onto observations, the r1w1 model bolometric light curve nonetheless closely matches the pseudo-bolometric light curve of SN 2023axu. This agreement further supports r1w1 as the preferred model for SN 2023axu.

Combining our analysis with the model light curve, we note that the strength of CSM interaction is tied to the mass-loss history of the progenitor star. Strong stellar winds (e.g., models r1w4, r1w6) produce a dense, extended CSM, leading to vigorous interaction that efficiently thermalizes kinetic energy into radiation—manifested as early brightening, linear declines (IIL-like), and flash spectral features. SN 2023axu, instead, exemplifies the weak-wind scenario (e.g., r1w1): its low-density CSM leads to very weak interaction, minimal conversion of kinetic energy to radiation, and a light curve that closely follows that of a normal SN IIP.

\subsection{Progenitor Mass}
\label{subsec:prog}
The nebular-phase spectrum of SN 2023axu ($t \sim$ 288 d) allows us to estimate the properties of its progenitor star. A common approach is to use the flux of the [O~I] $\lambda\lambda$ 6300,6364 doublet as a proxy for the CO core mass, which in turn reflects the progenitor mass \citep{10.1093/mnras/stu221, Uomoto1986}. However, the absolute flux is sensitive to uncertainties in distance and extinction. More robust indicators involve using the relative flux of the [O I] doublet, either normalized to the 5000–8500 \AA\,continuum \citep{fang2025redsupergiantproblemviewed} or compared to the [Ca II] $\lambda\lambda$7291, 7324 doublet \citep{Fransson1989, fang2022}, as these ratios are less sensitive to systematic uncertainties.

\begin{figure}[]
\includegraphics[width=8.4cm,angle=0]{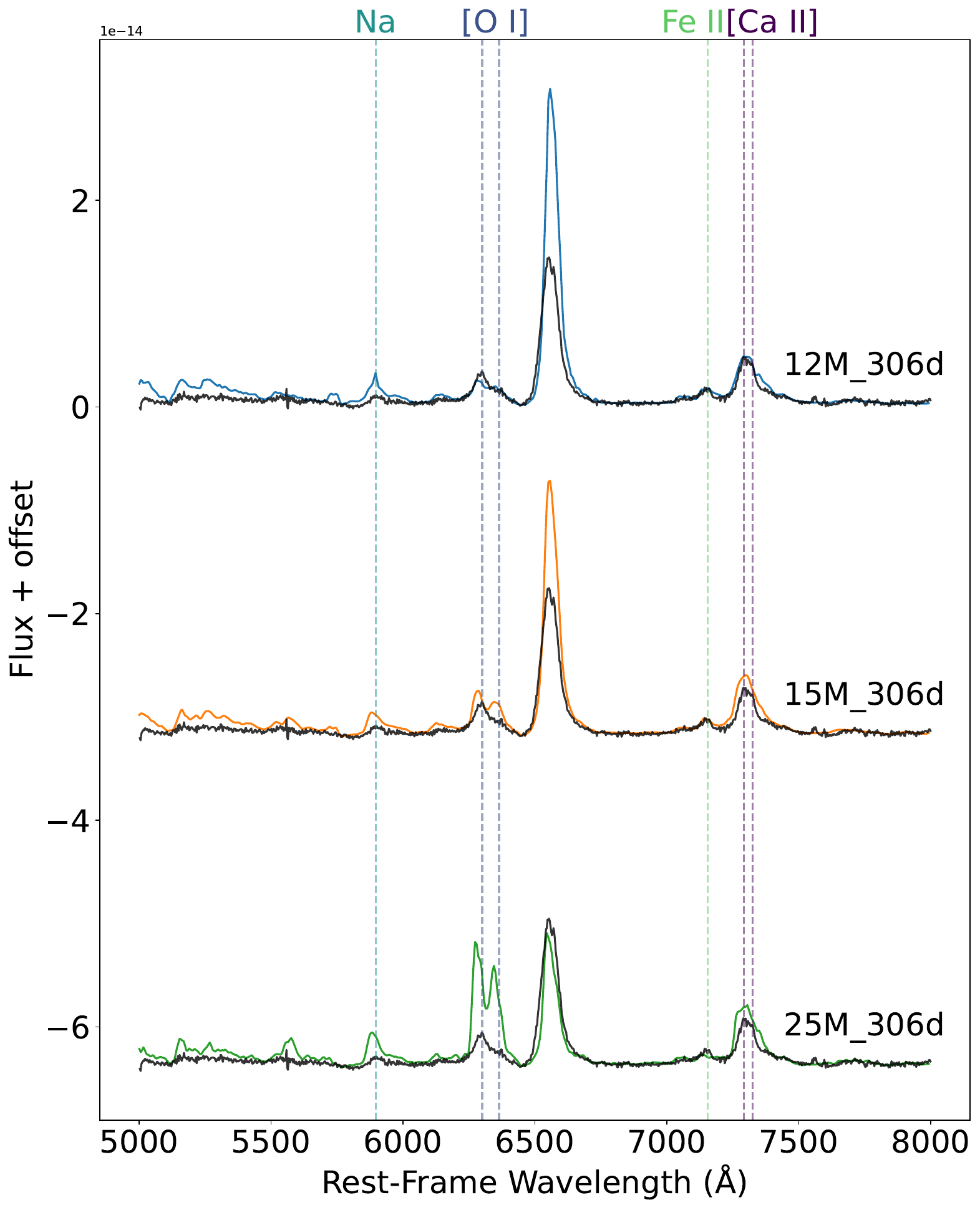}
\caption{Comparison of nebular spectra of SN 2023axu at $t \approx 288$ d with the 306 d model spectrum from \cite{2012A&A...546A..28J, 10.1093/mnras/stu221}. All model spectra were scaled to match the distance of SN 2023axu. The black line in the figure represents the nebular spectrum of SN~2023axu, while the spectra in other colors are model spectra for different progenitor masses and spectral phases.}
\label{fig:progenitor}
\end{figure}

\begin{figure*}
\centering
\includegraphics[width=0.45\textwidth]{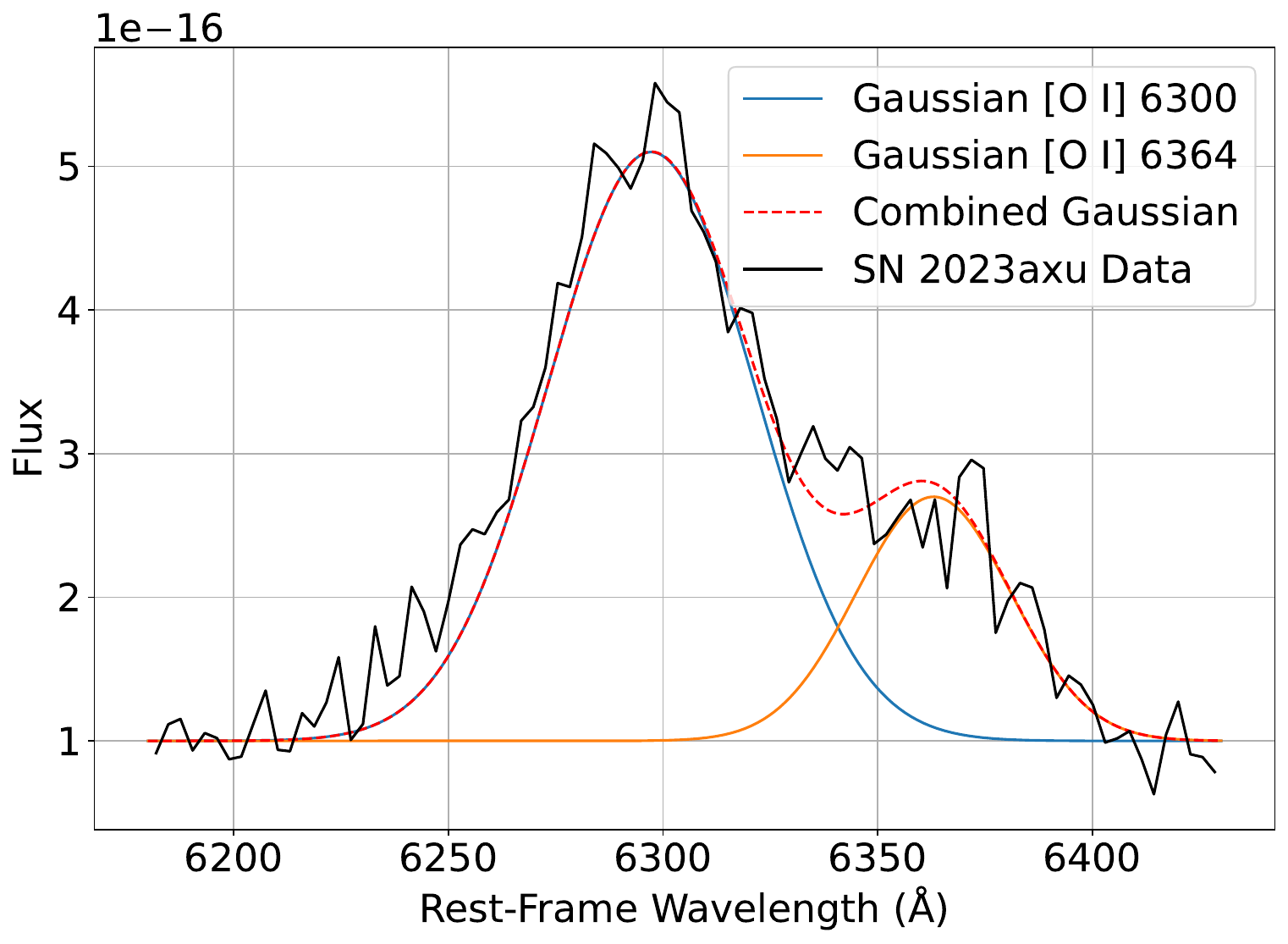}
\includegraphics[width=0.45\textwidth]{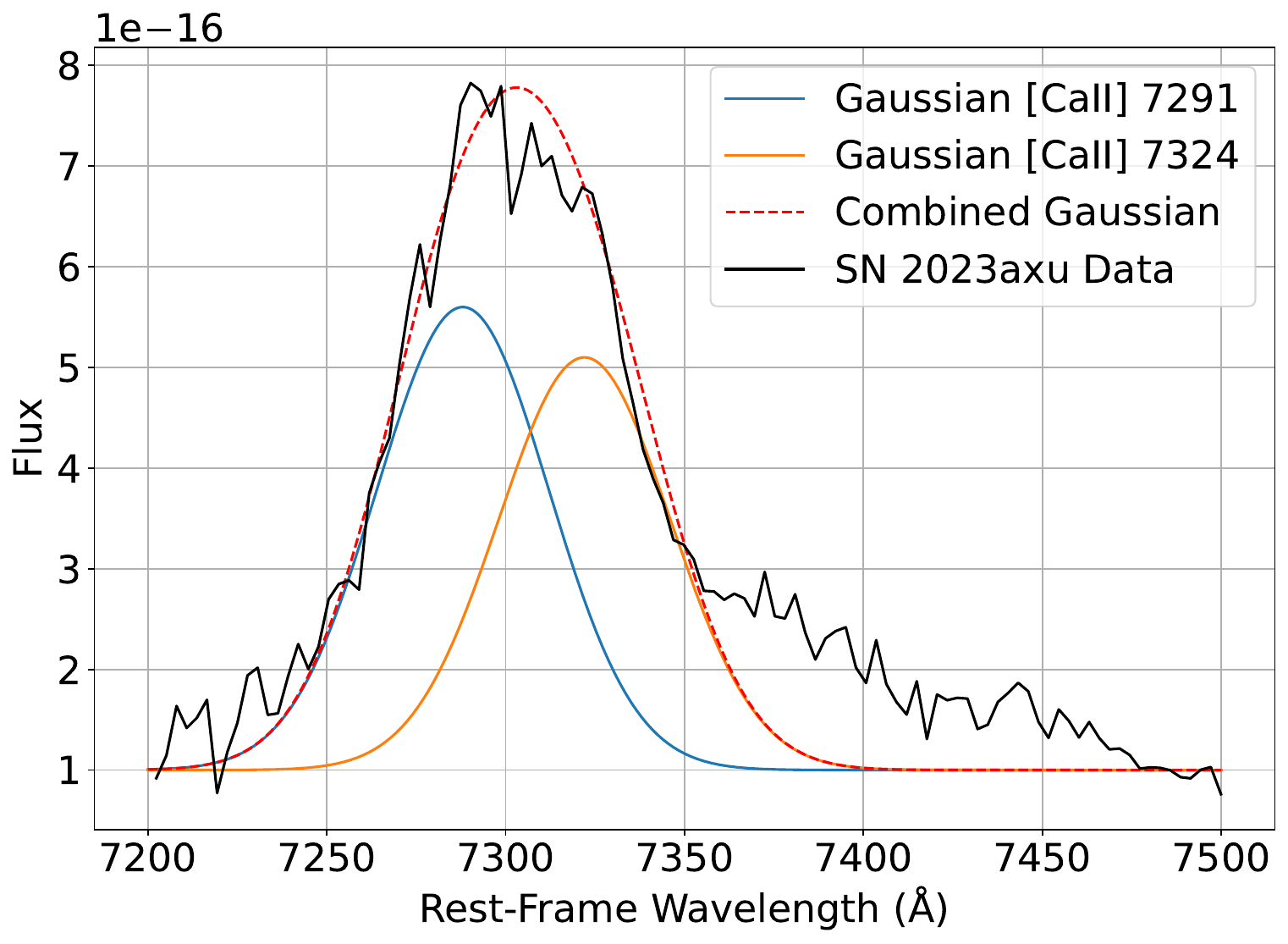}
\caption{Gaussian fitting of the [O I] $\lambda\lambda$6300, 6364 lines and [Ca II] $\lambda\lambda$7291, 7324 lines in the nebular spectrum of SN~2023axu.
\label{fig:gas}}
\end{figure*}

We compare the nebular spectrum of SN 2023axu with theoretical models from \citet{2012A&A...546A..28J, 10.1093/mnras/stu221} in Figure \ref{fig:progenitor}, which link spectral features to progenitor masses under the assumption of a retained hydrogen-rich envelope. In the figure, we overplot the 306 d model because it is closer in phase to our observation than the 250 d model and provides a notably better match to the continuum. The observed H$\alpha$ emission is significantly weaker than predicted by these models, this discrepancy likely results from substantial mass loss before explosion: hydrodynamic simulations show that SNe II with stripped hydrogen envelopes (M$_{Henv} < 3\, M_\odot$) exhibit suppressed H$\alpha$ flux due to lower recombination efficiency, while other spectral regions are less affected \citep{Dessartluc2022,fang2025redsupergiantproblemviewed}. Observations support this trend — SNe II showing plateau light curves indicative of low M$_{Henv}$ consistently display faint nebular H${\alpha}$ lines \citep{Fang_2025b,Teja_2022,Teja_2023,Ravi_2025}. Although CSM interaction could enhance H${\alpha}$ emission, spectroscopic analysis of SN 2023axu suggests minimal CSM interaction, ruling out this mechanism as a significant contributor \citep{DessartLuc2023}.

From the comparison in Figure \ref{fig:progenitor}, the 12 M$_\odot$ and 15~M$_\odot$ progenitor models at 306 d post-explosion show the best agreement with the observed flux ratio of [O I] to [Ca II]. We performed Gaussian fitting on the nebular spectrum (Figure \ref{fig:gas}) to measure these fluxes, obtaining a ratio of 0.61 for SN 2023axu. The same method applied to the 12 M$_\odot$ and 15 M$_\odot$ models at 306 d yields ratios of 0.49 and 0.68, respectively, supporting a progenitor mass between $12 - 15$ M$_\odot$.

We also applied the diagnostic of \citet{fang2025redsupergiantproblemviewed}, which uses the ratio of the [O I] $\lambda$$\lambda$ 6300,6364 doublet flux to the integrated flux within 5000--8500 \AA\,as a progenitor-mass indicator. To mitigate potential contamination, we subtracted the H$\alpha$contribution from the total. The derived relative flux for SN 2023axu is 0.194. This value is significantly higher than that of the 12 M$_\odot$ model at 306 d (0.104) and closely matches the 15 M$_\odot$ model at the same epoch (0.198), further supporting a progenitor mass closer to 15 M$_\odot$. We note that this result carries uncertainties, primarily from continuum flux estimation. Furthermore, all calculations rule out a high-mass progenitor at $M_{ZMAS} \geq 19\, M_\odot$.

We also noted that the Na $\lambda\lambda$5890,5896 line emission of SN 2023axu is significantly weaker than in the model. Although sodium nucleosynthesis does not vary monotonically with progenitor mass, this may also indicate hydrogen envelope stripping. In contrast, the Fe II $\lambda$7155 line intensity matches the models, suggesting that sodium exists in a highly ionized state. Given the low temperatures in the nebular phase, this could be due to a reduced recombination rate caused by very low density in the late-time ejecta. 

\section{CONCLUSIONS}
\label{sec:conclusion}
SN 2023axu reached a peak absolute $V-$ band magnitude of $M_{\rm V}= -17.25\pm0.06$ at about 7 d after the explosion, and produced about 0.055 $M_\odot$ of $^{56}$Ni. Combining constraints from nebular spectral modeling and emission-line ratios, we estimate a progenitor mass about 15 \Msun.

The spectral evolution of SN 2023axu is largely consistent with that of typical SNe IIP, though it shows a distinct broad ``ledge" feature near 4600 \AA\ in early phases. We have investigated the origin of this feature by comparing SN 2023axu with other SNe and with the spectral models of \citet{DessartLuc}, the early-time spectra are best reproduced by the weak-wind model (r1w1). The presence of additional broad emission features in the 0.67 d spectrum, along with the P-Cygni-like morphology of the ledge, supports the interpretation by \citet{10.1093/mnras/staf893} that the ledge is a P-Cygni profile formed in the ionized ejecta and sparse mass loss material remains confined near the SN. In the case of SN 2023axu, the feature likely arises from a blend of C, N, and He lines, rather than He alone. Furthermore, the late-time photometric decline rate approaches the $^{56}$Co decay rate, and the absence of discernible interaction power indicates no strong late-time interaction, consistent with the weak-wind scenario.

Comparison with well-studied SNe shows that the light curve of SN 2023axu is consistent with those of normal SNe IIP. SNe with early flash-ionized emission lines, such as SN 2018zd, typically exhibit steeper post-peak declines, resembling the IIL subtype. This divergence can be understood within the framework of \citet{DessartLuc}, which links the mass-loss history of the progenitor to the observed light-curve shape. Progenitors with high mass-loss rates develop dense CSM; the resulting strong ejecta–CSM interaction efficiently converts kinetic energy into radiation, leading to rapidly declining light curves and flash-ionized spectral features. SN~2023axu, on the other hand, exemplifies the weak-wind scenario: its tenuous CSM leads to minimal interaction, little additional thermalization of kinetic energy, and a light curve closely matching those of typical SNe~IIP.

In summary, the multi-wavelength analysis of SN~2023axu contributes to our understanding of the ledge feature and exemplifies the weak mass-loss pathway that leads to minimal CSM interaction, thereby helping to explain the observed divergence in light-curve evolution between SNe IIP and IIL. 

\section*{Acknowledgments}
This work is supported by the B-type Strategic Priority Program of the Chinese Academy of Sciences (Grant No. XDB1160202), the National Key R\&D Program of China with grant 2021YFA1600404, the National Natural Science Foundation of China (NSFC grants 12173082 and 12333008), the Yunnan Fundamental Research Projects (YFRP; grants 202501AV070012 and 202401BC070007), the Top-notch Young Talents Program of Yunnan Province, the Light of West China Program provided by the Chinese Academy of Sciences, and the International Centre of Supernovae (ICESUN), Yunnan Key Laboratory of Supernova Research (No. 202505AV340004). AP, AR, GV acknowledge support from the PRIN-INAF 2022, ``Shedding light on the nature of gap transients: from the observations to the models". AR also acknowledges financial support from the GRAWITA Large Program Grant (PI P. D'Avanzo). Yongzhi Cai is supported by the National Natural Science Foundation of China (NSFC, Grant No. 12303054), the National Key Research and Development Program of China (Grant No. 2024YFA1611603), the Yunnan Fundamental Research Projects (Grant Nos. 202401AU070063, 202501AS070078). We acknowledge the support of the staff of the LJT. Funding for the LJT has been provided by the CAS and the People's Government of Yunnan Province. The LJT is jointly operated and administrated by YNAO and Center for Astronomical Mega-Science, CAS. This article is also based on observations made with the Rapid Eye Mount Telescope, operated in La Silla by INAF under the program A47TAC\_37 (PI: G.~Valerin). Based on observations collected at Schmidt 67/92 telescope (Asiago Mount Ekar, Italy) of the INAF - Osservatorio Astronomico di Padova. 

\textbf{Facilities:} YAO:2.4m (LJT), WMS, REM, Schmidt 67/92 telescopes. 

\textbf{Software:} PyRAF \citep{2012ascl.soft07011S}, NumPy \citep{2020Natur.585..357H}, Matplotlib \citep{4160265}.

\bibliography{2023axu}{}
\bibliographystyle{aasjournalv7}

\appendix
\section{Photometric and spectroscopic data}
This section presents the observational data for SN 2023axu. The local photometric standard stars in the field of SN 2023axu are provided in Table \ref{tab:finder}. The photometric data from the LJT, WMS, REM, and Schmidt telescopes are provided in Table \ref{tab:A1}, ZTF and ATLAS photometric data are presented in Table \ref{tab:A2}, HST photometric data are presented in Table \ref{Tab:HST}, while the log of spectroscopic observations obtained with the LJT, including epoch, exposure time, and configuration, is given in Table~\ref{Tab:Spec_log}.

\begin{table*}[ht!]
\tablenum{A1}
\caption{Local photometric standard stars in the field of SN 2023axu}
\begin{tabular}{lccccccc}
\hline\hline
Star & B (mag)& V (mag)& g (mag)& r (mag)& i (mag)& z (mag)& H(mag) \\
\hline
1&16.177(0.09)&14.653(0.08)&15.30(0.01)&14.25(0.01)&13.79(0.01)&13.46(0.01)&11.611(0.03)\\
2&14.838(0.06)&14.192(0.09)&14.44(0.01)&13.98(0.01)&13.83(0.01)&13.75(0.01)&12.541(0.03)\\
3&18.983(0.09)&17.737(0.05)&16.97(0.01)&16.26(0.01)&15.97(0.01)&15.85(0.02)&15.284(0.14)\\
4&16.236(0.04)&14.859(0.13)&15.73(0.01)&14.30(0.01)&13.70(0.01)&13.28(0.01)&11.059(0.03)\\
5&15.798(0.02)&14.727(0.13)&15.34(0.01)&14.41(0.01)&14.05(0.01)&13.82(0.01)&12.137(0.02)\\
6&17.095(0.15)&16.183(0.13)&16.71(0.01)&16.01(0.01)&15.77(0.01)&15.68(0.02)&14.21(0.05)\\
7&14.44(0.02)&13.756(0.08)&13.18(0.01)&12.43(0.01)&12.10(0.01)&12.28(0.02)&12.123(0.02)\\
8&15.185(0.04)&14.522(0.07)&14.80(0.01)&14.28(0.01)&14.12(0.01)&14.04(0.01)&12.802(0.03)\\
9&14.618(0.03)&13.962(0.14)&14.23(0.01)&13.74(0.01)&14.12(0.02)&13.51(0.01)&12.363(0.02)\\
\hline
\hline
\end{tabular}

\label{tab:finder}
\end{table*}

\setcounter{table}{0}
\tablenum{A2}
\begin{longtable}{cccccccccc} 
\caption[A2]{Journal of photometric observations of SN 2023axu.}\label{tab:A1}\\ 
\midrule  
\endfirsthead  

\multicolumn{10}{c}  
{{\bfseries Table \thetable{} -- Continued}} \\  
\toprule MJD&Epoch (day)$^a$&B (mag)&V (mag)&g (mag)&r (mag)&i (mag)&z (mag)&H (mag)&Facility\\
\midrule  
\endhead

\midrule  
\multicolumn{10}{c}{} \\  
\endfoot  

\bottomrule  
\endlastfoot

MJD&Epoch (day)$^a$&B (mag)&V (mag)&g (mag)&r (mag)&i (mag)&z (mag)&H (mag)&Facility\\
\hline
59972.68 	&	1.18	&...&...&	15.23(0.04)&	15.25(0.02)&	15.30(0.05)&...&...&	LJT	\\
59974.71 	&	3.21	&	15.14(0.01)&	14.82(0.01)&	14.83(0.05)&	14.67(0.04)&	14.70(0.05)&...&...&	LJT	\\
59976.70 	&	5.2	&	15.08(0.01)&	14.71(0.01)&	14.76(0.06)&	14.53(0.04)&	14.49(0.04)&...&...&	LJT	\\
59977.69 	&	6.19	&	15.04(0.01)&	14.67(0.01)&	14.71(0.05)&	14.48(0.11)&	14.24(0.35)&...&...&	LJT	\\
59979.64 	&	8.14	&	15.04(0.02)&	14.66(0.01)&	14.70(0.07)&	14.41(0.05)&	14.41(0.05)&...&...&	LJT	\\
59981.58 	&	10.08	&	15.07(0.02)&	14.68(0.01)&	14.74(0.06)&	14.45(0.06)&	14.46(0.03)&...&...&	LJT	\\
59982.74 	&	11.24	&	15.12(0.02)&	14.70(0.01)&	14.76(0.04)&	14.41(0.04)&	14.45(0.05)&...&...&	LJT	\\
59985.61 	&	14.11	&	15.14(0.02)&	14.72(0.01)&	14.83(0.06)&	14.45(0.06)&	14.49(0.03)&...&...&	LJT	\\
59989.52 	&	18.02	&...&...&	14.86(0.04)&	14.41(0.05)&	14.41(0.06)&...&...&	LJT	\\
59990.60 	&	19.09	&...&...&	14.81(0.07)&	14.49(0.06)&	14.42(0.08)&...&...&	WMS	\\
59990.64 	&	19.14	&	15.31(0.01)&	14.74(0.01)&	14.90(0.05)&	14.44(0.06)&	14.45(0.09)&...&...&	LJT	\\
59991.59 	&	20.08	&...&...&	15.01(0.06)&	14.41(0.09)&	14.36(0.09)&...&...&	WMS	\\
59992.59 	&	21.09	&...&...&	15.00(0.05)&	14.45(0.06)&	14.25(0.04)&...&...&	WMS	\\
59993.57 	&	22.06	&...&...&	15.01(0.08)&	14.43(0.05)&	14.19(0.05)&...&...&	WMS	\\
59994.59 	&	23.08	&...&...&	15.03(0.05)&	14.43(0.04)&	14.39(0.07)&...&...&	WMS	\\
59995.57 	&	24.06	&...&...&	15.00(0.07)&	14.40(0.09)&	14.25(0.04)&...&...&	WMS	\\
59996.56 	&	25.06	&...&...&	15.06(0.05)&	14.42(0.06)&	14.18(0.09)&...&...&	WMS	\\
59998.58 	&	27.07	&...&...&	15.16(0.08)&	14.41(0.06)&	14.41(0.07)&...&...&	WMS	\\
60000.57 	&	29.06	&...&...&	15.13(0.05)&	14.45(0.05)&	14.38(0.10)&...&...&	WMS	\\
60001.55 	&	30.05	&...&...&	15.15(0.03)&	14.49(0.08)&	14.33(0.08)&...&...&	WMS	\\
60002.55 	&	31.05	&...&...&	15.16(0.07)&	14.48(0.06)&	14.33(0.07)&...&...&	WMS	\\
60003.56 	&	32.06	&...&...&	15.20(0.08)&	14.53(0.07)&	14.40(0.07)&...&...&	WMS	\\
60004.53 	&	33.03	&	15.72(0.01)&	14.87(0.01)&	15.15(0.06)&	14.47(0.08)&	14.45(0.05)&	14.33(0.04)&...&	LJT	\\
60006.78  &  35.28  &  15.85(0.03) &  14.90(0.02) &  …  &  …  &  …  &  …  &  …  &  Schmidt\\
60010.53 	&	39.03	&	15.82(0.01)&					&	15.2(0.05)&	14.48(0.07)&	14.44(0.08)&...&...&	LJT	\\
60015.55 	&	44.05	&	15.93(0.01)&	14.94(0.01)&	15.28(0.02)&	14.52(0.03)&...&...&...&	LJT	\\
60027.54 	&	56.04	&	16.12(0.01)&	15.05(0.01)&	15.43(0.06)&	14.65(0.08)&	14.54(0.02)&	14.40(0.03)&...&	LJT	\\
60028.03 	&	56.53	&...&...&	15.39(0.06)&	14.22(0.05)&	14.12(0.06)&	14.43(0.07)&	13.02 (0.06)&	rem	\\
60032.04 	&	60.54	&...&...&	15.37(0.04)&	14.63(0.03)&	14.48(0.05)&	14.41(0.07)&	13.13 (0.06)&	rem	\\
60037.01 	&	65.51	&...&...&	15.52(0.06)&	14.63(0.04)&	14.52(0.04)&	14.41(0.08)&	13.17 (0.08)&	rem	\\
60041.02 	&	69.52	&...&...&	15.47(0.06)&	14.70(0.03)&	14.59(0.08)&	14.52(0.08)&	13.16 (0.08)&	rem	\\
60041.53 	&	70.03	&	16.31(0.01)&	15.18(0.01)&	15.57(0.04)&	14.69(0.08)&	14.64(0.07)&	14.43(0.05)&...&	LJT	\\
60045.02 	&	73.52	&...&...&	15.54(0.05)&	14.71(0.07)&	14.64(0.06)&	14.54(0.07)&	13.19 (0.07)&	rem	\\
60046.53 	&	75.03	&	16.35(0.01)&	15.22(0.01)&...&	14.72(0.05)&	14.65(0.06)&	14.52(0.07)&...&	LJT	\\
60049.02 	&	77.52	&...&...&	15.52(0.06)&	14.70(0.06)&	14.66(0.08)&	14.47(0.08)&	13.20 (0.06)&	rem	\\
60049.53 	&	78.03	&	16.23(0.09)&	15.27(0.06)&...&	14.76(0.10)&	14.77(0.12)&	14.44(0.08)&...&	LJT	\\
60052.53 	&	81.03	&	16.46(0.02)&	15.31(0.01)&	15.78(0.03)&	14.85(0.04)&	14.78(0.03)&	14.58(0.04)&...&	LJT	\\
60053.02 	&	81.52	&...&...&	15.66(0.05)&	14.92(0.11)&	14.76(0.09)&	14.60(0.07)&	13.22 (0.07)&	rem	\\
60057.05 	&	85.55	&...&...&	15.79(0.05)&	14.89(0.03)&	14.80(0.04)&	14.62(0.06)&	13.31 (0.08)&	rem	\\
60058.53 	&	87.03	&	16.59(0.03)&	15.44(0.01)&	15.93(0.04)&	14.95(0.05)&	14.88(0.02)&	14.68(0.04)&...&	LJT	\\
60061.04 	&	89.54	&...&...&	15.82(0.08)&	15.01(0.12)&	14.90(0.08)&	14.71(0.10)&	13.39 (0.06)&	rem	\\
60065.05 	&	93.55	&...&...&	15.97(0.13)&	14.96(0.08)&	14.71(0.07)&	14.82(0.10)&	13.51 (0.07)&	rem	\\
60069.05 	&	97.55	&...&...&	16.66(0.19)&	15.30(0.07)&	15.28(0.10)&	15.18(0.12)&	13.69 (0.07)&	rem	\\
60073.05 	&	101.55	&...&...&	17.20(0.22)&	15.55(0.05)&	15.48(0.09)&	15.32(0.11)&	13.83 (0.06)&	rem	\\
60077.06 	&	105.56	&...&...&	17.14(0.15)&	15.95(0.14)&	16.06(0.21)&	15.56(0.11)&	13.93 (0.08)&	rem	\\
60239.89 	&	268.39	&...&	18.66(0.03)&	19.09(0.09)&	17.53(0.07)&	17.90(0.08)&...&...&	LJT	\\
60255.82 	&	284.32	&...&...&...&	17.73(0.06)&	17.95(0.42)&...&...&	LJT	\\
60270.82 	&	299.32	&...&	18.96(0.05)&	19.26(0.11)&	17.91(0.05)&	18.24(0.07)&	17.78(0.08)&...&	LJT	\\
60368.51 	&	397.01	&...&...&...&	19.49(0.16)&	19.41(0.15)&	19.12(0.15)&...&	LJT	\\
60370.56 	&	399.06	&...&...&...&	19.36(0.11)&	19.17(0.13)&...&...&	LJT	\\
60382.54 	&	411.04	&...&...&...&	19.26(0.34)&	19.12(0.17)&...&...&	LJT	\\
\hline
\hline

\end{longtable}

\setcounter{table}{0}
\tablenum{A3}
\begin{longtable}{cccccccccc} 
\caption[A3]{ATLAS and ZTF observations of SN 2023axu.}\label{tab:A2}\\ 
\midrule  
\endfirsthead  

\multicolumn{5}{c}  
{{\bfseries Table \thetable{} -- Continued}} \\  
\toprule MJD	&Epoch (day)&	MAG(err)&filter	&	TELESCOPE	\\

\midrule  
\endhead

\midrule  
\multicolumn{5}{c}{} \\  
\endfoot  

\bottomrule  
\endlastfoot 
MJD	&Epoch (day)&	MAG(err)&	filter	&	TELESCOPE	\\
\hline
59971.91 	&	0.41 	&	15.93(0.01)&c	&	ATLAS	\\
59972.27 	&	0.77 	&	15.47(0.03)&r	&	ZTF	\\
59973.13 	&	1.63 	&	14.95(0.01)&o	&	ATLAS	\\
59974.54 	&	3.04 	&	14.59(0.01)&o	&	ATLAS	\\
59975.87 	&	4.37 	&	14.51(0.01)&o	&	ATLAS	\\
59977.15 	&	5.65 	&	14.42(0.01)&o	&	ATLAS	\\
59978.51 	&	7.01 	&	14.39(0.01)&o	&	ATLAS	\\
59979.21 	&	7.71 	&	14.40(0.03)&r	&	ZTF	\\
59979.92 	&	8.42 	&	14.39(0.01)&o	&	ATLAS	\\
59981.15 	&	9.65 	&	14.38(0.01)&o	&	ATLAS	\\
59982.17 	&	10.67 	&	14.32(0.01)&o	&	ATLAS	\\
59982.23 	&	10.73 	&	14.38(0.02)&r	&	ZTF	\\
59983.54 	&	12.04 	&	14.36(0.01)&o	&	ATLAS	\\
59984.90 	&	13.40 	&	14.41(0.01)&o	&	ATLAS	\\
59985.21 	&	13.71 	&	14.38(0.02)&r	&	ZTF	\\
59985.88 	&	14.38 	&	14.40(0.01)&o	&	ATLAS	\\
59987.15 	&	15.65 	&	14.40(0.01)&o	&	ATLAS	\\
59988.89 	&	17.39 	&	14.37(0.01)&o	&	ATLAS	\\
59989.89 	&	18.39 	&	14.36(0.01)&o	&	ATLAS	\\
59991.14 	&	19.64 	&	14.70(0.01)&c	&	ATLAS	\\
59991.19 	&	19.69 	&	14.84(0.02)&g	&	ZTF	\\
59991.21 	&	19.71 	&	14.31(0.02)&r	&	ZTF	\\
59993.84 	&	22.34 	&	14.36(0.01)&o	&	ATLAS	\\
59995.08 	&	23.58 	&	14.75(0.01)&c	&	ATLAS	\\
59997.89 	&	26.39 	&	14.35(0.01)&o	&	ATLAS	\\
60003.07 	&	31.57 	&	14.40(0.01)&o	&	ATLAS	\\
60005.86 	&	34.36 	&	14.40(0.01)&o	&	ATLAS	\\
60007.11 	&	35.61 	&	14.41(0.01)&o	&	ATLAS	\\
60011.48 	&	39.98 	&	14.44(0.01)&o	&	ATLAS	\\
60013.23 	&	41.73 	&	14.42(0.02)&r	&	ZTF	\\
60013.25 	&	41.75 	&	15.24(0.03)&g	&	ZTF	\\
60013.89 	&	42.39 	&	14.47(0.01)&o	&	ATLAS	\\
60015.14 	&	43.64 	&	14.45(0.01)&o	&	ATLAS	\\
60017.18 	&	45.68 	&	15.19(0.05)&g	&	ZTF	\\
60017.81 	&	46.31 	&	14.99(0.16)&c	&	ATLAS	\\
60019.06 	&	47.56 	&	14.45(0.01)&o	&	ATLAS	\\
60021.15 	&	49.65 	&	14.49(0.02)&r	&	ZTF	\\
60021.84 	&	50.34 	&	15.03(0.01)&c	&	ATLAS	\\
60023.08 	&	51.58 	&	14.46(0.01)&o	&	ATLAS	\\
60025.78 	&	54.28 	&	15.07(0.02)&c	&	ATLAS	\\
60027.05 	&	55.55 	&	14.53(0.01)&o	&	ATLAS	\\
60029.83 	&	58.33 	&	15.10(0.01)&c	&	ATLAS	\\
60030.15 	&	58.65 	&	15.44(0.03)&g	&	ZTF	\\
60031.11 	&	59.61 	&	14.55(0.01)&o	&	ATLAS	\\
60033.77 	&	62.27 	&	14.55(0.01)&o	&	ATLAS	\\
60035.82 	&	64.32 	&	14.56(0.01)&o	&	ATLAS	\\
60036.15 	&	64.65 	&	15.54(0.03)&g	&	ZTF	\\
60036.17 	&	64.67 	&	14.58(0.03)&r	&	ZTF	\\
60037.03 	&	65.53 	&	14.56(0.01)&o	&	ATLAS	\\
60038.83 	&	67.33 	&	14.58(0.01)&o	&	ATLAS	\\
60040.08 	&	68.58 	&	14.62(0.01)&o	&	ATLAS	\\
60041.82 	&	70.32 	&	14.62(0.01)&o	&	ATLAS	\\
60043.08 	&	71.58 	&	14.63(0.01)&o	&	ATLAS	\\
60044.37 	&	72.87 	&	14.68(0.01)&o	&	ATLAS	\\
60046.05 	&	74.55 	&	15.29(0.01)&c	&	ATLAS	\\
60047.78 	&	76.28 	&	14.71(0.02)&o	&	ATLAS	\\
60049.07 	&	77.57 	&	15.35(0.01)&c	&	ATLAS	\\
60050.81 	&	79.31 	&	14.72(0.01)&o	&	ATLAS	\\
60052.08 	&	80.58 	&	15.40(0.01)&c	&	ATLAS	\\
60053.80 	&	82.30 	&	14.79(0.01)&o	&	ATLAS	\\
60056.06 	&	84.56 	&	15.49(0.01)&c	&	ATLAS	\\
60058.06 	&	86.56 	&	15.54(0.01)&c	&	ATLAS	\\
60062.36 	&	90.86 	&	14.98(0.09)&o	&	ATLAS	\\
60065.81 	&	94.31 	&	15.09(0.02)&o	&	ATLAS	\\
60067.04 	&	95.54 	&	15.15(0.01)&o	&	ATLAS	\\
60068.38 	&	96.88 	&	15.23(0.02)&o	&	ATLAS	\\
60070.74 	&	99.24 	&	15.34(0.01)&o	&	ATLAS	\\
60072.02 	&	100.52 	&	15.37(0.02)&o	&	ATLAS	\\
60074.03 	&	102.53 	&	15.56(0.01)&o	&	ATLAS	\\
60074.74 	&	103.24 	&	16.47(0.02)&c	&	ATLAS	\\
60080.74 	&	109.24 	&	16.88(0.03)&c	&	ATLAS	\\
60083.76 	&	112.26 	&	17.00(0.03)&c	&	ATLAS	\\
60084.98 	&	113.48 	&	16.18(0.02)&o	&	ATLAS	\\
60085.00 	&	113.50 	&	16.14(0.02)&o	&	ATLAS	\\
60092.73 	&	121.23 	&	16.21(0.03)&o	&	ATLAS	\\
60180.14 	&	208.64 	&	16.98(0.05)&o	&	ATLAS	\\
60183.15 	&	211.65 	&	17.00(0.03)&o	&	ATLAS	\\
60187.34 	&	215.84 	&	17.06(0.06)&o	&	ATLAS	\\
60189.13 	&	217.63 	&	17.08(0.06)&o	&	ATLAS	\\
60204.39 	&	232.89 	&	17.16(0.03)&o	&	ATLAS	\\
60210.11 	&	238.61 	&	18.26(0.06)&c	&	ATLAS	\\
60211.34 	&	239.84 	&	17.30(0.04)&o	&	ATLAS	\\
60214.34 	&	242.84 	&	17.38(0.05)&o	&	ATLAS	\\
60215.48 	&	243.98 	&	18.82(0.16)&g	&	ZTF	\\
60216.01 	&	244.51 	&	17.20(0.10)&o	&	ATLAS	\\
60217.34 	&	245.84 	&	17.22(0.06)&o	&	ATLAS	\\
60219.09 	&	247.59 	&	17.40(0.06)&o	&	ATLAS	\\
60221.12 	&	249.62 	&	17.48(0.07)&o	&	ATLAS	\\
60222.38 	&	250.88 	&	17.34(0.05)&o	&	ATLAS	\\
60222.46 	&	250.96 	&	18.85(0.16)&g	&	ZTF	\\
60222.50 	&	251.00 	&	17.29(0.03)&r	&	ZTF	\\
60224.46 	&	252.96 	&	18.91(0.13)&g	&	ZTF	\\
60224.50 	&	253.00 	&	17.35(0.04)&r	&	ZTF	\\
60227.97 	&	256.47 	&	17.49(0.06)&o	&	ATLAS	\\
60228.02 	&	256.52 	&	17.56(0.04)&o	&	ATLAS	\\
60228.43 	&	256.93 	&	18.84(0.14)&g	&	ZTF	\\
60229.31 	&	257.81 	&	17.48(0.04)&o	&	ATLAS	\\
60231.45 	&	259.95 	&	18.96(0.13)&g	&	ZTF	\\
60231.50 	&	260.00 	&	17.45(0.04)&r	&	ZTF	\\
60234.06 	&	262.56 	&	17.54(0.04)&o	&	ATLAS	\\
60235.49 	&	263.99 	&	18.90(0.12)&g	&	ZTF	\\
60237.46 	&	265.96 	&	18.95(0.12)&g	&	ZTF	\\
60237.50 	&	266.00 	&	17.48(0.04)&r	&	ZTF	\\
60239.44 	&	267.94 	&	17.51(0.04)&r	&	ZTF	\\
60239.48 	&	267.98 	&	19.00(0.10)&g	&	ZTF	\\
60240.30 	&	268.80 	&	17.65(0.05)&o	&	ATLAS	\\
60241.29 	&	269.79 	&	17.65(0.08)&o	&	ATLAS	\\
60242.04 	&	270.54 	&	17.69(0.06)&o	&	ATLAS	\\
60243.09 	&	271.59 	&	17.59(0.07)&o	&	ATLAS	\\
60244.30 	&	272.80 	&	17.42(0.13)&o	&	ATLAS	\\
60244.40 	&	272.90 	&	19.10(0.26)&g	&	ZTF	\\
60246.50 	&	275.00 	&	17.64(0.05)&r	&	ZTF	\\
60248.00 	&	276.50 	&	17.81(0.11)&o	&	ATLAS	\\
60248.01 	&	276.51 	&	17.79(0.11)&o	&	ATLAS	\\
60249.05 	&	277.55 	&	17.76(0.09)&o	&	ATLAS	\\
60249.39 	&	277.89 	&	19.07(0.23)&g	&	ZTF	\\
60249.47 	&	277.97 	&	17.76(0.14)&r	&	ZTF	\\
60251.36 	&	279.86 	&	18.94(0.17)&g	&	ZTF	\\
60251.43 	&	279.93 	&	17.67(0.06)&r	&	ZTF	\\
60253.30 	&	281.80 	&	17.59(0.05)&o	&	ATLAS	\\
60253.38 	&	281.88 	&	17.69(0.05)&r	&	ZTF	\\
60253.45 	&	281.95 	&	19.22(0.16)&g	&	ZTF	\\
60254.46 	&	282.96 	&	19.10(0.13)&g	&	ZTF	\\
60254.50 	&	283.00 	&	17.73(0.05)&r	&	ZTF	\\
60255.28 	&	283.78 	&	17.77(0.05)&o	&	ATLAS	\\
60256.27 	&	284.77 	&	17.84(0.05)&o	&	ATLAS	\\
60258.38 	&	286.88 	&	19.12(0.19)&g	&	ZTF	\\
60258.44 	&	286.94 	&	17.77(0.04)&r	&	ZTF	\\
60259.99 	&	288.49 	&	18.81(0.10)&c	&	ATLAS	\\
60260.05 	&	288.55 	&	18.82(0.09)&c	&	ATLAS	\\
60260.35 	&	288.85 	&	19.26(0.07)&g	&	ZTF	\\
60260.37 	&	288.87 	&	17.78(0.05)&r	&	ZTF	\\
60262.27 	&	290.77 	&	17.73(0.05)&o	&	ATLAS	\\
60262.47 	&	290.97 	&	17.83(0.05)&r	&	ZTF	\\
60262.54 	&	291.04 	&	19.17(0.13)&g	&	ZTF	\\
60264.28 	&	292.78 	&	17.94(0.06)&o	&	ATLAS	\\
60266.07 	&	294.57 	&	18.91(0.10)&c	&	ATLAS	\\
60268.25 	&	296.75 	&	17.95(0.06)&o	&	ATLAS	\\
60270.23 	&	298.73 	&	18.08(0.08)&o	&	ATLAS	\\
60279.00 	&	307.50 	&	18.11(0.09)&o	&	ATLAS	\\
60279.00 	&	307.50 	&	18.08(0.09)&o	&	ATLAS	\\
60280.99 	&	309.49 	&	18.06(0.07)&o	&	ATLAS	\\
60282.34 	&	310.84 	&	18.10(0.04)&o	&	ATLAS	\\
60286.93 	&	315.43 	&	18.22(0.08)&o	&	ATLAS	\\
60287.29 	&	315.79 	&	19.53(0.25)&g	&	ZTF	\\
60288.40 	&	316.90 	&	18.20(0.08)&r	&	ZTF	\\
60288.40 	&	316.90 	&	18.22(0.03)&r	&	ZTF	\\
60288.44 	&	316.94 	&	19.46(0.16)&g	&	ZTF	\\
60288.44 	&	316.94 	&	19.59(0.12)&g	&	ZTF	\\
60289.30 	&	317.80 	&	18.22(0.03)&o	&	ATLAS	\\
60289.34 	&	317.84 	&	18.14(0.07)&r	&	ZTF	\\
60290.24 	&	318.74 	&	18.23(0.04)&o	&	ATLAS	\\
60290.37 	&	318.87 	&	19.48(0.16)&g	&	ZTF	\\
60290.37 	&	318.87 	&	19.46(0.09)&g	&	ZTF	\\
60290.38 	&	318.88 	&	18.15(0.04)&r	&	ZTF	\\
60290.38 	&	318.88 	&	18.13(0.02)&r	&	ZTF	\\
60291.02 	&	319.52 	&	18.25(0.07)&o	&	ATLAS	\\
60291.35 	&	319.85 	&	18.17(0.06)&r	&	ZTF	\\
60292.02 	&	320.52 	&	18.29(0.07)&o	&	ATLAS	\\
60293.31 	&	321.81 	&	18.02(0.07)&o	&	ATLAS	\\
60294.71 	&	323.21 	&	18.23(0.07)&o	&	ATLAS	\\
60295.96 	&	324.46 	&	18.24(0.07)&o	&	ATLAS	\\
60297.21 	&	325.71 	&	18.29(0.10)&o	&	ATLAS	\\
60298.22 	&	326.72 	&	18.29(0.08)&o	&	ATLAS	\\
60299.03 	&	327.53 	&	18.40(0.08)&o	&	ATLAS	\\
60306.36 	&	334.86 	&	18.38(0.14)&r	&	ZTF	\\
60308.38 	&	336.88 	&	18.33(0.26)&r	&	ZTF	\\
60314.28 	&	342.78 	&	19.64(0.23)&g	&	ZTF	\\
60314.28 	&	342.78 	&	19.68(0.12)&g	&	ZTF	\\
60314.42 	&	342.92 	&	18.45(0.09)&r	&	ZTF	\\
60314.42 	&	342.92 	&	18.47(0.05)&r	&	ZTF	\\
60322.25 	&	350.75 	&	18.73(0.14)&r	&	ZTF	\\
60322.27 	&	350.77 	&	19.92(0.25)&g	&	ZTF	\\
60324.22 	&	352.72 	&	19.62(0.22)&g	&	ZTF	\\
60324.24 	&	352.74 	&	18.72(0.09)&r	&	ZTF	\\
60324.24 	&	352.74 	&	18.74(0.06)&r	&	ZTF	\\
60326.23 	&	354.73 	&	18.68(0.07)&r	&	ZTF	\\
60326.32 	&	354.82 	&	19.76(0.22)&g	&	ZTF	\\
60326.32 	&	354.82 	&	19.76(0.10)&g	&	ZTF	\\
60334.21 	&	362.71 	&	18.89(0.25)&r	&	ZTF	\\
60339.23 	&	367.73 	&	18.85(0.11)&r	&	ZTF	\\
60339.23 	&	367.73 	&	18.89(0.07)&r	&	ZTF	\\
60339.36 	&	367.86 	&	20.20(0.30)&g	&	ZTF	\\
60352.24 	&	380.74 	&	19.06(0.11)&r	&	ZTF	\\
60368.29 	&	396.79 	&	19.35(0.22)&r	&	ZTF	\\
60368.29 	&	396.79 	&	19.42(0.19)&r	&	ZTF	\\
60388.15 	&	416.65 	&	21.06(0.66)&g	&	ZTF	\\
60390.15 	&	418.65 	&	19.58(0.20)&r	&	ZTF	\\

\hline
\hline

\end{longtable}

\begin{table*}[ht!]
\tablenum{A4}
\caption{Photometric observations of SN 2023axu by HST}
\begin{tabular}{ccccccc}
\hline\hline
MJD & Epoch (day)$^a$ & F814W(mag) & F555W(mag) & F438W(mag) & F336W(mag) & F275W(mag) \\
\hline
60942.8 & 971.3 & 25.26(0.12) & 26.27(0.15) & 26.55(0.50) & $>$25.58 & $>$24.58 \\
\hline
\hline
\end{tabular}

$^a${All of the data presented in this article were obtained from the MAST at the Space Telescope Science Institute. The specific observations analyzed can be accessed via \dataset[doi:10.17909/6djx-c648]{https://doi.org/10.17909/6djx-c648}.}
\label{Tab:HST}
\end{table*}

\begin{table*}[ht!]
\tablenum{A5}
\caption{Journal of spectroscopic observations of SN~2023axu by LJT with YFOSC}
\begin{tabular}{lcccccc}
\hline\hline
Date (UT) & MJD & Epoch (day)$^a$ & Range (\AA) & Disp. (\AA\ pix$^{-1}$) & Exp (s) & airmass \\
\hline
2023-01-28 & 59972.67 & 1.17 & 3600-8950 & 2.85 & 900 & 1.43 \\
2023-01-30 & 59974.70 & 3.20 & 3600-8950 & 2.85 & 700 & 1.53 \\
2023-02-01 & 59976.68 & 5.18 & 3600-8950 & 2.85 & 900 & 1.50 \\
2023-02-04 & 59979.63 & 8.13 & 3600-8950 & 2.85 & 700 & 1.41 \\
2023-02-06 & 59981.56 & 10.06 & 3600-8950 & 2.85 & 700 & 1.53 \\
2023-02-07 & 59982.73 & 11.23 & 3600-8950 & 2.85 & 700 & 1.93 \\
2023-02-12 & 59987.52 & 16.02 & 3600-8950 & 2.85 & 1200 & 1.69 \\
2023-02-15 & 59990.62 & 19.12 & 3600-8950 & 2.85 & 1000 & 1.43 \\
2023-02-24 & 59999.58 & 28.08 & 3600-8950 & 2.85 & 800 & 1.41 \\
2023-03-01 & 60004.52 & 33.02 & 3600-8950 & 2.85 & 1000 & 1.47 \\
2023-03-07 & 60010.53 & 39.03 & 3600-8950 & 2.85 & 1000 & 1.41 \\
2023-03-12 & 60015.53 & 44.03 & 3600-8950 & 2.85 & 1000 & 1.41 \\
2023-03-29 & 60032.53 & 61.03 & 3600-8950 & 2.85 & 1000 & 1.49 \\
2023-04-06 & 60040.53 & 69.03 & 3600-8950 & 2.85 & 1000 & 1.62 \\
2023-04-12 & 60046.54 & 75.04 & 3600-8950 & 2.85 & 1000 & 1.81 \\
2023-11-11 & 60259.83 & 288.33 & 3600-8950 & 2.85 & 2800 & 1.44 \\
\hline
\hline
\end{tabular}

$^a${The epoch is relative to the explosion date, MJD = 59971.5.}
\label{Tab:Spec_log}
\end{table*}

\end{CJK}
\end{document}